\documentclass[sn-mathphys,Numbered]{sn-jnl}
\usepackage{graphicx}%
\usepackage{multirow}%
\usepackage{amsmath,amssymb,amsfonts}%
\usepackage{amsthm}%
\usepackage{mathrsfs}%
\usepackage[title]{appendix}%
\usepackage{xcolor}%
\usepackage{textcomp}%
\usepackage{manyfoot}%
\usepackage{booktabs}%
\usepackage{algorithm}%
\usepackage{algorithmicx}%
\usepackage{algpseudocode}%
\usepackage{listings}%
\usepackage{soul}%
\usepackage{subcaption}%
\usepackage[justification=centering, singlelinecheck=false]{caption}
\theoremstyle{thmstyleone}%
\theoremstyle{thmstyletwo}%

\theoremstyle{thmstylethree}%

\begin{document}

\title[Article Title]{Transferable 3D Convolutional Neural Networks for Elastic Constants Prediction in Nanoporous Metals}
\author*[1,2]{\fnm{Sergei} \sur{Zorkaltsev}}\equalcont{These authors contributed equally to this work.}\email{sergei.zorkaltsev@imdea.org}
\author[3]{\fnm{Rafał} \sur{Topolnicki}}\equalcont{These authors contributed equally to this work.}
\author[4]{\fnm{Tal-El} \sur{Carmon}}\equalcont{These authors contributed equally to this work.}
\author[4]{\fnm{Santhosh} \sur{Mathesan}}
\author[3]{\fnm{Paweł} \sur{Dłotko}}
\author*[4]{\fnm{Dan} \sur{Mordehai}}\email{danmord@me.technion.ac.il}
\author*[2]{\fnm{Maciej} \sur{Haranczyk}}\email{maciej.haranczyk@imdea.org}

\affil[1]{\orgname{Universidad Carlos III de Madrid}, \orgaddress{\street{Av. Universidad 30}, \city{Leganes, Madrid}, \postcode{28911}, \country{Spain}}}

\affil[2]{\orgname{IMDEA Materials Institute}, \orgaddress{\street{C. Eric Kandel 2}, \city{Getafe, Madrid}, \postcode{28906}, \country{Spain}}}

\affil[3]{\orgname{Institute of Mathematics, Polish Academy of Sciences}, \orgaddress{\street{ul. Śniadeckich 8}, \city{Warsaw}, \postcode{00-656}, \country{Poland}}}

\affil[4]{\orgname{Faculty of Mechanical Engineering, Technion-Israel Institute of Technology}, \orgaddress{\city{Haifa}, \postcode{3200003}, \country{Israel}}}

\abstract{The topology of nanoporous metals is crucial for determining their mechanical response. In this work, we generated 6,000 gold and 422 silver nanoporous structures and calculated three components of elastic modulus with Molecular Dynamics simulations, resulting in 19,263 data points. This study compared two distinct approaches of predicting elastic modulus: a Fully-Connected neural network trained on precomputed topological descriptors, and several 3D Convolutional neural network architectures adapted from computer vision. The 3D CNNs outperformed the descriptor‐based baseline model ($R^2 = 0.704$), with top-performing DenseNet‐201 architecture achieving $R^2 = 0.955$. Additionally, the effects of training grid resolution, dataset size, and descriptor integration into a model were investigated. We further demonstrated model robustness through Transfer learning: a pretrained model was fine-tuned on a much smaller dataset of denser gold structures and the dataset of denser silver structures. Finally, the trained model was employed to evaluate the mechanical properties of 100,000 stochastic nanoporous gold structures and identify the Pareto optimal designs.
}


\maketitle

\section{Introduction}\label{Intro}

Nanoporous metals (such as noble: Au, Ag, Pd, Pt and non-noble: Al, Ni, Cu, Zn) have found a wide range of applications in many fields: chemical catalysis \cite{catalysis1,catalysis2,catalysis3}, sensors \cite{sensor1, sensor2}, and energy storage \cite{storage1, storage2} because of their high conductivity, high surface-to-volume ratio, and catalytic activity. Depending on desired mechanical properties, porosity, and application, these nanoporous structures can be manufactured by de-alloying \cite{dealloying1, dealloying2}, chemical deposition \cite{deposition}, template fabrication \cite{templating}, and other methods. Existing approaches to determining the relationship between the microstructure and mechanical properties of nanoporous metals are mostly based on scaling laws. One of these scaling laws is the Gibson-Ashby model \cite{gibson}, which correlates the elastic modulus and the yield stress of the structure with its solid fraction, or the Roberts-Garboczi model \cite{RGmodel_2002}, which corrects the results for small solid fractions considering a percolation threshold.

Although accurate in predicting properties of some types of materials, these scaling laws can deviate substantially for nanoporous metals \cite{winter}, lattices \cite{zhong2023low}, and triply periodic minimal surface (TPMS) structures \cite{tripleperiodic}. For instance, in the study of Peng et al. \cite{silver_improved}, researchers manufactured nanoporous silver samples by joining silver nanowires, reaching porous structure with high solid fractions (above 50\%), for which the Gibson-Ashby model is believed to be accurate. However, they found out that the yield strength of nanosilver porous structure is higher than predicted by the Gibson-Ashby model, reaching 2.6 GPa. To explain this, they have shown that the yield strength also depends on the thickness and length of the nanowires, obeying a smaller-is-stronger law. On the other hand, experimental studies have shown that the Gibson–Ashby scaling law remains valid for nanoporous structures when size effects and network connectivity or topological parameters are appropriately considered in predicting their mechanical properties \cite{Liu_network, Hiang_network}.

In addition to experimental approaches, simulations enable a more detailed analysis of topology and its relationship to mechanical properties. These studies facilitate the examination and modification of the existing scaling laws by incorporating new topologically related parameters or proposing novel equations. Some finite element method (FEM) studies pointed out that the effective Young's modulus of nanoporous structures also depends on the scaled genus, proposing either a linear \cite{mangipudi_topology-dependent_2016, mangipudi_morphological_2017} or a quadratic \cite{sohn_scaling_2024} relationship. The precision of topology-independent scaling laws deteriorates in nanoporous structures in confined volume, as was demonstrated in nanoporous gold (NPG) pillars \cite{wu_consequences_2023}, as the number of ligaments that support the deformation is limited. The need to account for the load-bearing ligaments was previously demonstrated in molecular dynamics (MD) simulations \cite{Mathesan}. In these compression simulations, it was shown that the number of load-bearing ligaments can differentiate the Young's moduli of NPG nanopillars of the same solid fraction but with different topologies. Accordingly, MD simulations can serve as a powerful tool to find structure-property correlations in nanoporous structures, since elastic properties are well captured by interatomic potentials and the simulation cells usually include only a few load-bearing ligaments, which makes the results more sensitive to topology. Despite this, the relations between the structures and elastic properties of nanoporous stochastic structures remain unresolved.

As an alternative to experiments and simulations, deep learning (DL) has demonstrated great success in many research areas, and it can be used to predict properties directly on input structures. The popularity of DL models comes from a specific type of neural network called Convolutional Neural Networks (CNNs). Initially developed for 2D data and particularly Computer Vision related problems, the architectures like AlexNet \cite{alexnet}, VGG \cite{vgg}, ResNet \cite{resnet}, and their modifications are still applied for many tasks, including materials science. For instance, Kato et al. \cite{steel_cnn} trained a CNN to classify microstructure images of samples produced from two different alloys with two different heat treatment conditions. Their results show that with only 140 images available for training, it is possible to achieve near-ideal classification accuracy with data augmentation. In the work of Wu et al. \cite{permeability_cnn} authors developed a similar CNN-based approach for the regression task of permeability prediction. The neural network architecture of this study was a physics-informed CNN, which takes both 2D images and physical parameters (porosity and surface area ratio) as the inputs. This approach allowed for achieving high accuracy of predictions while being orders of magnitude faster than conventional methods for permeability estimation.

It is also possible to apply convolutions to three-dimensional data by 3D-CNNs. Cawte and Bazylak \cite{gas_3dcnn} trained a 3D-CNN to predict the permeability of numerically generated gas diffusion layer materials. A relatively simple and straightforward CNN architecture of this work showed a low prediction error with a mean 5-fold $R^{2}$ score of 0.9885. The authors concluded that the performance of this simple deep learning model is comparable to the best available models for permeability prediction. Rao and Liu \cite{homogenization_3dcnn} applied 3D-CNN to a dataset of generated composite materials with random spherical inclusions. The mechanical properties of these materials were estimated with FEM simulations and used as targets for model training. Their results showed a high accuracy of 3D-CNN predictions, lower computational costs compared to conventional methods, and the potential to transfer learning, which allows training on a smaller dataset. Santos et al. \cite{flow_3dcnn} developed a multiscale 3D-CNN model to predict fluid flow through a porous media. This multi-scale approach combined several subsequent networks, each of them being aware of previous predictions and having a finer size of convolutional filter. The model was trained on a mix of synthetic and real data and then tested on vuggy core samples, realistic sandstone samples, and micro-CT images of fractured carbonates. Other examples of tasks solved with the application of 3D-CNN include natural convection in porous materials \cite{convection_3dcnn}, up-scaling of rock micro-CT images \cite{upscale_3dcnn}, detection of clogging in microfluidic systems \cite{clogging_3dcnn}, and prediction of effective thermal conductivity \cite{conductivity_3dcnn}. 

As discussed above, DL techniques can be used to uncover structure–property relationships in a variety of material systems. In the context of the present study, the relationship between nanoporous structures and their mechanical properties has been specifically examined \cite{CNN_tomography, ML_wang_biomaterial, Huber_dimension, shargh_interpretable_2023, Niaz_NP, Chen_Huber_2025, Dennis_AI}. CNN architectures were successfully employed in the generation of realistic synthetic training nanoporous structures either from experimental focused ion beam (FIB) tomography images \cite{CNN_tomography} or from physics-informed phase-field simulations \cite{ML_wang_biomaterial}, which could open pathways for further studies previously limited by insufficient training data. The mechanical properties of the nanoporous structures used in the prediction of the structure-property relationships were primarily quantified from FEM simulations \cite{Niaz_NP, Huber_dimension, Dennis_AI} or from FFT mechanical model \cite{ML_wang_biomaterial}. These synthetic data and the stiffness information were further employed in the design of nanoporous structures to achieve the target anisotropic stiffness tensor, with predominant biomedical applications such as dental and hip implants \cite{ML_wang_biomaterial}. In addition to the well-known effect of solid volume fraction on the stiffness of nanoporous structures, CNN could be an efficient tool to identify unexplored design parameters and their contributions. For instance, the saddle-shaped regions of the ligaments were identified to decrease the effective stiffness of nanoporous structures \cite{Niaz_NP}, which was previously unknown. In addition to  CNN architecture, the DL framework could enable the inverse design of nanoporous structures by fine-tuning the design input parameters in the Gaussian random field \cite{Dennis_AI} that could generate the synthetic data, enabling the additive manufacturing of these structures with target anisotropy information. In summary, DL approaches demonstrate significant potential for the tailored design of nanoporous structures \cite{Chen_Huber_2025}, based on the input from experiments or physics-based simulations. Recent studies have also applied Convolutional Recurrent Neural Networks (CRNN) to accelerate microstructure evolution and other image-based materials problems \cite{lanzoni2022morphological, yang2021self, lanzoni2024extreme}. They demonstrated that 3D convolutional architectures in combination with recurrent or physics-informed layers can learn spatiotemporal patterns and predict microstructural and morphological evolution over time. There is existing literature focusing on RVEs and atomistic data to predict mechanical responses and stiffness tensors. For instance, Eidel \cite{eidel2023deep} demonstrated 3DCNN models that predict full stiffness tensors and stiffness bounds for a wide range of phase properties and boundary conditions. Aldakheel et al. \cite{aldakheel2023efficient} studied efficient 2DCNN architectures with transfer learning for heterogeneous composites. Mianroodi et al. \cite{mianroodi2022lossless} implemented image-based models to recover atomistic effects in nanoporous aluminium (obtained from 2D representation). Peivaste et al. \cite{ peivaste2024rapid} demonstrated that 3DCNNs capture elastic responses caused by the structural defect at the atomistic scale. In contrast, the present work trains and interprets fully 3D Convolutional networks on realistic and complex nanoporous metal samples. A large MD-labelled dataset demonstrates robustness to voxel resolution, is efficient in cross-material and porosity transfer learning, and practical large-scale screening.

In this study, we explored a deep learning approach for predicting a mechanical property of nanoporous metals. First, an extensive database with a total of 6,422 nanoporous structures was generated, including 6,000 gold and 422 silver structures with two different target solid fractions. Molecular Dynamics (MD) simulations were used to calculate the elastic modulus in three principal directions for the entire database, resulting in 19,266 datapoints. We evaluated two distinct approaches to establish the structure-property relationship in nanoporous metals: one based on feature extraction with convolutions, and the other based on directly providing features as a calculated set of descriptors. Several common CNN architectures were adapted to work with 3D data, and their performance was compared with each other and with a fully-connected neural network trained on topological descriptors. We demonstrated the flexibility and robustness of the trained models through Transfer learning by reusing previously learned patterns to refine the models with a much smaller amount of training data. Finally, the trained model was used to rapidly evaluate the elastic modulus of 100,000 randomly generated gold structures and identify Pareto optimal ones.

\section{Methods}\label{Methods}

\subsection {Data generation}
\subsubsection{Nanoporous metal structures}\label{structure generation}

\begin{figure*}[h]
    \centering
    \includegraphics [width=1.0\textwidth]{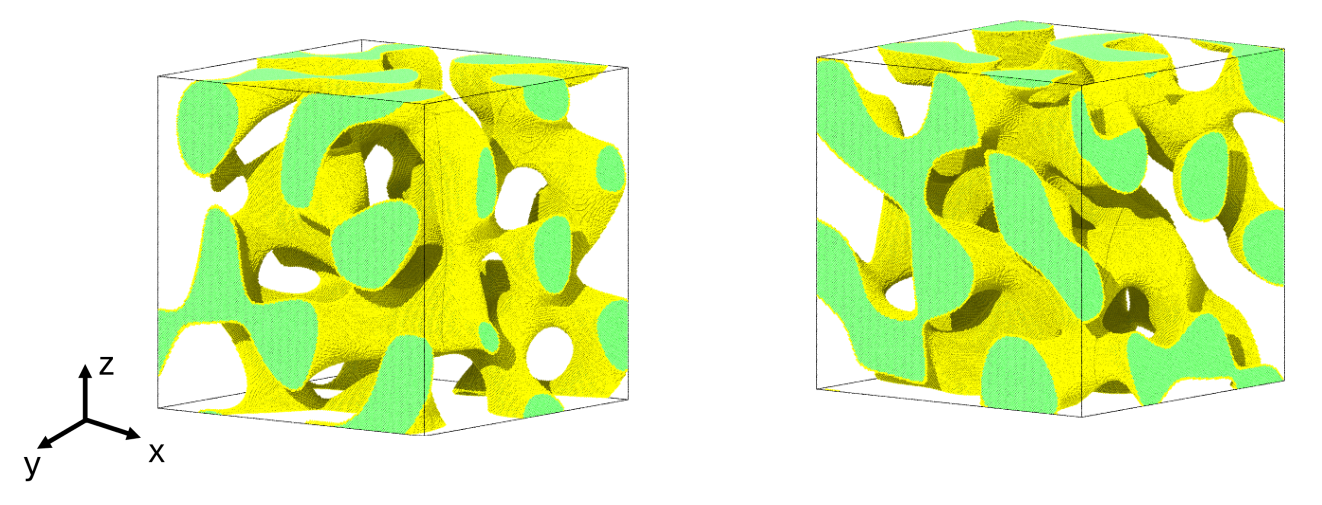}
    \caption{Typical NPG structures with solid volume fraction of 0.25 (left) and 0.35 (right). (Color: Yellow - surface atoms and Green - FCC atoms).}
  \label{NPG_Figure_1}
\end{figure*}

Three-dimensional periodic bicontinuous nanoporous structures were generated based on the method proposed by Soyarslan et al. \cite{SOYARSLAN}, which involves superposition of standing sinusoidal waves with fixed wavelengths and varying phases. Within this method, the face-centered cubic (FCC) lattice basis was adopted to generate single-crystalline FCC nanoporous structures, with $\langle 100 \rangle$ orientations aligned parallel to the cubic simulation box. The size of the edges was equal to $125$ unit cells in each direction. This simulation cell allows, on the one hand, to generate a large number of different topologies, but at the same time, to obtain a large distribution of mechanical properties due to the few load-bearing ligaments in each configuration. The parameters used to generate the nanoporous structures are detailed in \textit{Supplementary Information, Sec.1}. A total of 6,422 nanoporous gold (NPG) and silver structures with two desired solid volume fractions: 0.25 and 0.35, were generated and visualized in OVITO  \cite{ovito}, as shown in Figure \ref{NPG_Figure_1}.

\subsubsection{Quantitative descriptors of nanoporous structures}

Nanoporous structures were characterized by quantifying the morphological and topological parameters, commonly referred to as descriptors. Morphological descriptors are related to the shape, size, and geometric features of the nanoporous structures, and they are quantified using 4 descriptors, namely solid volume fraction, node diameter, ligament diameter, and ligament curvature. Topological descriptors are related to network connectivity and are defined from 26 descriptors, namely, genus, number of nodes, directional, and volumetric resistance of elements (both ligaments and nodes) in nanoporous structures with respect to the loading axes $x$, $y$, and $z$. 

Genus is defined as the number of closed loops within the connected network as observed in nanoporous structures \cite{Mathesancms, Richert}. The number of nodes refers to the total number of intersection points at which three or more ligaments connect to form a three-dimensional connected network structure. Following the method described by Mathesan et al. \cite{Mathesan}, the structure was skeletonized and a directional descriptor, also referred to as the number of load-bearing ligaments, is quantified from the total projection of the skeleton elements onto the loading axis, divided by the length of the simulation box along the axis. Furthermore, instead of uniformly accounting for the directional contribution of all the skeleton elements, a modified directional descriptor is defined by assigning higher weights to elements that are more closely aligned with the applied loading direction. We name these descriptors as ligament-only directional resistance and modified ligament-only directional resistance, respectively (resulting in a total of six directional descriptors). 

However, the skeleton disregards the size of the nodes, which do not show any preferential orientation as opposed to ligaments. In this work, we account for six additional directional descriptors that take into account the predominantly isotropic characteristics of the nodes. We differentiate between the volume of each node and the ligaments, which are outside this volume. Each node is defined as the minimal sphere, centered at the skeleton node, that is confined within the volume of the nanoporous structure, and ligaments are now defined only as skeleton elements outside these spheres. The directional resistance descriptor is then defined as the total projection of the skeleton elements onto the loading axis and the diameter of the spheres,  divided by the length of the simulation box along the axis. Similarly, in the modified directional resistance descriptor, higher weights are assigned to elements that are more aligned with the loading direction.

Although the directional descriptors offer valuable insight into the load-bearing capacity of nanoporous structures, they do not account for variations in the cross-sectional area of individual elements, which is important in stochastic nanoporous structures. To incorporate both the directional alignment and the cross-sectional area, we define an additional 12 descriptors, similar to those described above, but we consider volumes instead of length. Instead of ligament length, we consider a volume measure, which is the length of the skeleton element multiplied by the area of the ligament's cross-section at the same point. When nodes are accounted for, the sphere's volume is considered instead of the diameter. Accordingly, we name these descriptors as ligament-only volumetric resistance and modified ligament-only volumetric resistance, when only skeleton elements are considered, and volumetric resistance and modified volumetric resistance, when nodes are accounted for. Altogether, the nanoporous structures are quantified from 30 descriptors (4 morphological + 26 topological) and they are summarized in Table \ref{descriptors}. The details of the methods involved in quantification of these descriptors are mentioned in the \textit{Supplementary Information, Sec.2}.

\begin{table}[h]
\caption{Summary of 30 descriptors to characterize the nanoporous structures}\label{tab2}%
\begin{tabular}{@{}ll@{}}
\toprule
Morphological & Topological \\
\midrule
Solid volume fraction & Genus \\
Node diameter & Number of nodes \\
Ligament diameter & Ligament-only directional resistance (LDR) ($x$, $y$, $z$) \\
Ligament curvature & Modified ligament-only directional resistance (MLDR) ($x$, $y$, $z$)\\
& Directional resistance (DR) ($x$, $y$, $z$) \\
& Modified directional resistance (MDR) ($x$, $y$, $z$) \\
& Ligament-only volumetric resistance (LVR) ($x$, $y$, $z$)\\
& Modified ligament-only volumetric resistance (MLVR) ($x$, $y$, $z$)\\
& Volumetric resistance (VR) ($x$, $y$, $z$)\\
& Modified volumetric resistance (MVR) ($x$, $y$, $z$)\\
\botrule
\end{tabular}
\label{descriptors}
\end{table}

\subsubsection{Molecular Dynamics simulations}

The dataset corresponding to the mechanical properties of nanoporous structures was quantified from a series of MD simulations performed using the open-source MD simulator LAMMPS \cite{LAMMPS}. The interactions between the individual gold (Au) or silver (Ag) atoms in the nanoporous structures were modeled with the embedded-atom method (EAM) interatomic potential proposed by Grochola et al. \cite{eam}. Periodic boundary conditions were applied in all three dimensions. The initial stress was minimized by relaxing the atomic positions using the conjugate gradient method. 

The elastic constants were quantified from static MD simulations, which consist of incremental deformation followed by energy minimization. The simulation box was strained by applying quasi-static compression followed by quasi-static tension to the original simulation box size. After each deformation step, the system was relaxed through energy minimization. The total applied strain ranged from ±2.5\%, applied in five discrete decrements and increments. This methodology was systematically repeated individually along the x, y, and z directions. Consequently, for each loading direction, there were 10 steps, and a linear fit was employed to describe the relation between the corresponding strained box dimensions and stress values. The fitting equation was given by $\sigma_{ii} = C_{ii}\frac{L}{L_0} - C_{ii}$, where  $\sigma_{ii}$ is the normal stress along the \(i^{\text{th}}\) axis where the simulation box was strained,  \(L\) is the instantaneous length of the box, \(L_0\) is the fitting parameter, \(C_{ii}\) is the elastic constant in the \(i^{\text{th}}\) axis, namely, $\ C_{11}, C_{22}$ and $\  C_{33} $ and \(i = 1, 2 \text{ or } 3\) correspond to axes $x$, $y$, and $z$, respectively,

\subsection{Artificial Neural Networks}
In this work, we implemented several CNN architectures common in Computer Vision tasks, modifying them and replacing 2D convolutional layers with 3D ones. Although each individual convolutional kernel extracts only local features, repeated application of convolutions across the neural network, combined with the activation function and pooling, progressively enlarges the receptive field. Therefore, deeper layers integrate information from large fractions of the structure, allowing the network to capture local ligament details and global connectivity patterns, determining the elastic response.
All the details of the implemented architectures and code can be found in the GitHub repository of this study.

\subsubsection{MobileNet}

MobileNets \cite{mobilenets} are a class of lightweight convolutional neural network architectures, designed as an alternative to deeper and more complicated ones. The authors of MobileNet demonstrated that the architecture performs on the same level of accuracy as much deeper models like GoogleNet, VGG, AlexNet, and others in common computer vision problems. This performance is achieved by factorizing standard convolution operations into depthwise separable convolutions followed by point-wise convolution. It was shown that factorization drastically reduces the model size and, therefore, the computational costs of the training. 

The network begins with a standard 3D convolution layer (3$\times$3$\times$3 kernel), followed by batch normalization and ReLU activation function. Next, data is passed through an array of blocks, each constructed by 3D depthwise convolution and pointwise convolution, both followed by batch normalization and an activation function. After the last block, the data is passed through the average pooling layer and flattened. Finally, the flattened feature maps are passed through a dropout and a fully-connected layer to output the model’s prediction.

\subsubsection{ResNet}
ResNets (Residual Neural Networks) \cite{resnet} were developed as a solution to common problems in extremely deep neural nets - vanishing/exploding gradients and degradation of training. ResNets solve these issues by introducing skip connections, which allow the gradient to reach all the layers of deep architectures during backpropagation. 

The model begins with a 3D convolution (7$\times$7$\times$7 kernel), followed by
batch normalization, ReLU activation, and max pooling layers. After the initial convolution, the model includes an array of residual blocks. Each of these blocks has a main convolution and a skip connection that bypasses it. The output of the convolution layer is summed with the skip connection before it is passed to ReLU activation function. In case there is a mismatch between input and output dimensions due to downsampling, the dimensions of the skip connection are adjusted by 1$\times$1$\times$1 convolution. After all convolution blocks, the model ends with adaptive average pooling and a fully connected layer, which outputs the final prediction.

\subsubsection{DenseNet}

DenseNets (Dense Convolutional Neural Networks) \cite{densenet} introduced a novel way of connecting convolutional layers. In DenseNets, each convolutional layer receives inputs from all the previous ones and passes its own output to all the subsequent layers. This architecture encourages feature reuse, increasing the input variation and improving efficiency.

The model starts with a 3D convolutional layer (7$\times$7$\times$7 kernel), followed by max pooling, and then proceeds with an array of dense blocks. Each dense block consists of several dense layers, which have the following structure: a 1$\times$1$\times$1 kernel convolution to reduce feature maps dimensions, followed by a 3$\times$3$\times$3 kernel convolution. Each dense layer concatenates its output with the input before passing the feature map to the next layer in the block. Transition layers connect dense blocks while reducing feature maps dimensions with 1$\times$1$\times$1 convolutions and average pooling.
After the final dense block, feature maps are passed through batch normalization, an activation function, and average pooling before returning the prediction from a fully connected layer.

\subsubsection{Fully-Connected NN}

Unlike the architectures mentioned before that operate on 2D or 3D data, Fully-Connected networks require a 1D feature vector as an input. In our case, the input vector was constructed from morphological and topological descriptors calculated for generated nanoporous structures. 
The architecture consists of 5 fully-connected layers with 1024, 512, 256, 128, and 64 neurons, respectively. Each layer, except the last one, is followed by batch normalization, ReLU activation, and dropout to prevent overfitting. The input vector is sequentially passed through all the layers, and the final layer outputs a single value, corresponding to a target property.

\subsection{Transfer learning}

During the training of a neural network, weights and bias terms of the model are iteratively adjusted to produce outputs that minimize a loss function. In the CNNs, these are the weights of filters in the convolutional layers and the weights of the fully-connected layers. Transfer learning is a technique that uses the knowledge gained by a model pretrained on one set of data to possibly improve the training on related but different data \cite{transfer}. Instead of starting the training with randomly initialized weights, transfer learning allows for reusing some or all of the weights from a pretrained model. This training strategy has an important advantage: the amount of data required for transfer learning is typically lower compared to the training of the initial model, which leads to a shorter time spent on training and lower computational costs.

Additionally, convolutional neural networks extract features hierarchically from training data \cite{high_low_features}. The initial layers detect low-level features such as horizontal or vertical lines, sharp edges, or other simple patterns. Deeper layers are able to detect high-level patterns, such as geometric shapes, and the last layers are activated on specific problem-related features. For example, in an image classification task, the last layer would detect the presence of an object of a specific class, e.g., a human, a dog, or a building. In the CNNs trained on nanoporous structures, the early layers capture local topological features from ligaments' geometry, while deeper layers extract more complex relationships between the whole ligaments' network and the target property. This suggests that the patterns learned on one set of structures can be reused to improve training on similar structures with a different base material or solid fraction.

\section{Results and discussion}\label{Results}

\subsection{Generated datasets}\label{sub:datasets}

\begin{figure}[h]
    \centering
    \begin{subfigure}[b]{0.4\textwidth}
        \includegraphics[width=\textwidth]{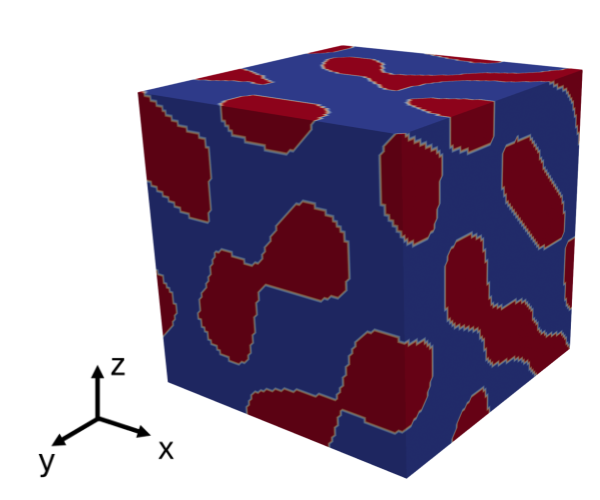}
        \caption{All values}
        \label{3d_all}
    \end{subfigure}
    \hfill
    \begin{subfigure}[b]{0.4\textwidth}
        \includegraphics[width=\textwidth]{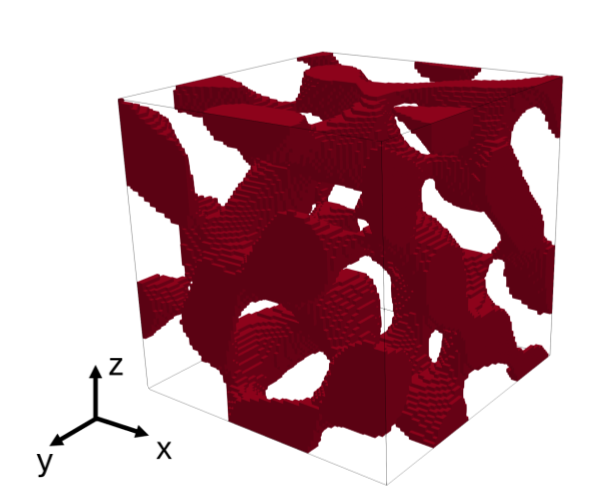}
        \caption{Solid part}
        \label{3d_solid}
    \end{subfigure}
    \caption{Nanoporous structure represented by a 3D binary array: (a) All values, (b) Solid part.}
    \label{3d_array}
\end{figure}

Nanoporous structures generated from the methodology described in Section \ref{structure generation} were initially saved in $.lmp$ format, which stores atom types, atom coordinates, and bounding box dimensions. Each data sample was converted into a 3D binary matrix (Figure \ref{3d_array}) to serve as an appropriate input data for CNNs training. We examine different voxel resolutions, ranging from  $30\times30\times30$ to  $80\times80\times80$. In this study, a total of 3 Datasets were defined to train and test the model, evaluate the performance of the model, and finally, implement the transfer learning. 

Dataset 1 includes 5000 NPG structures with a solid volume fraction of 0.25 and the corresponding MD-computed elastic constants, $\ C_{11}$,$\ C_{22}$, and $\ C_{33} $, resulting in a total of 15000 data points. In addition, a set of descriptors described in  Table \ref{descriptors} was calculated for each sample. In this way, Dataset 1 had two different representations (descriptors and binary matrix) of each nanoporous structure to train the models on and compare how the representation affects the model's ability to predict elastic constants. Dataset 1 was used to train and compare the performance of different CNN architectures, to evaluate the influence of training dataset size and binary grid resolution.

\begin{figure}[h]
    \centering
    \includegraphics [width=0.9\textwidth]{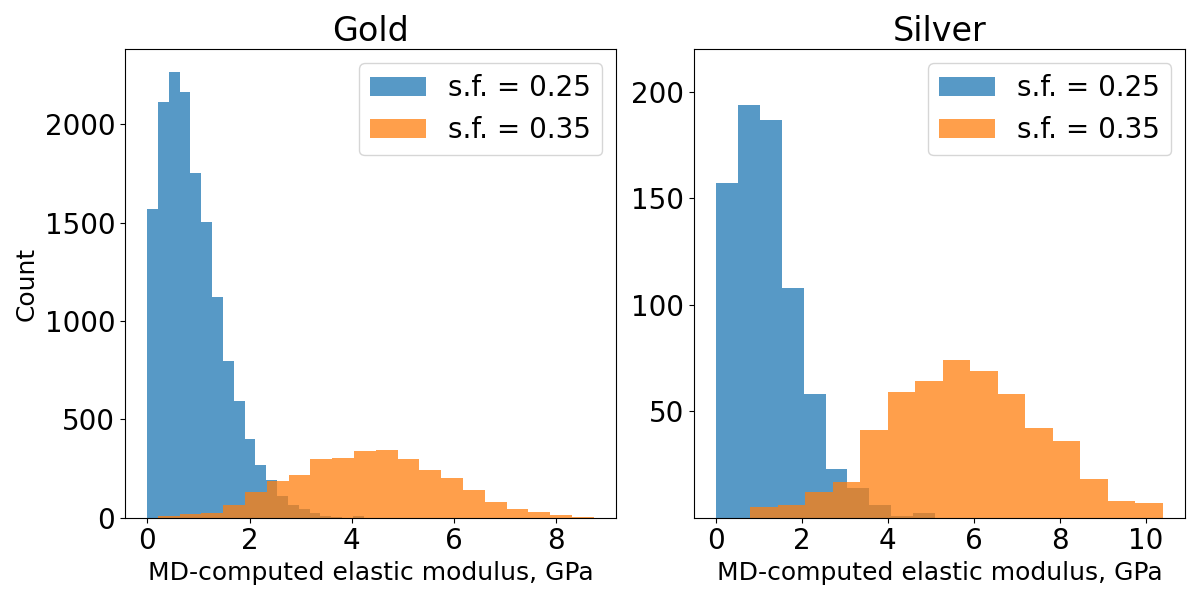}
    \caption{Distributions of MD-computed elastic constants $\ C_{ii}$ in the datasets. Each nanoporous structure is represented by three elastic constant values.}
  \label{computed_modulus}
\end{figure}

\begin{figure}[h]
    \centering
    \includegraphics [width=0.9\textwidth]{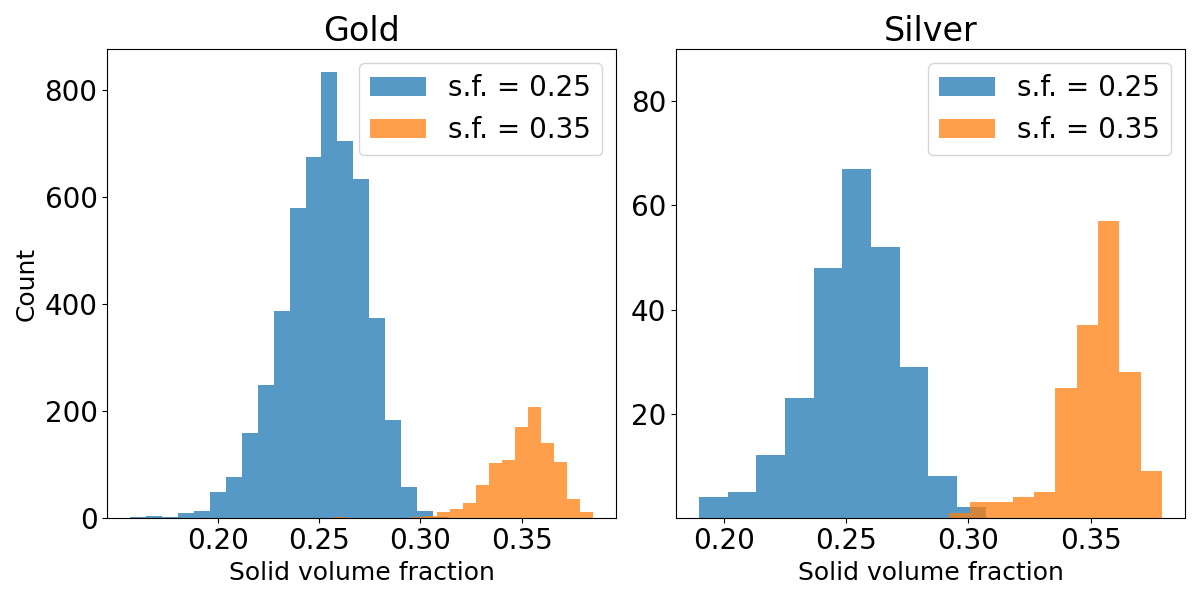}
    \caption{Distributions of as-built solid fraction in the datasets. Each nanoporous structure is represented by one solid fraction value.}
  \label{solid_fraction}
\end{figure}

Dataset 2 includes 1000 NPG structures with a solid volume fraction of 0.35 and the corresponding elastic constants (3000 data points). This dataset was used to evaluate the performance of the model trained on thinner structures from Dataset 1 and to implement transfer learning. Dataset 3 includes 422 nanoporous silver structures with MD-computed elastic constants. The target solid volume fractions of structures 1-250 and 251-422 were 0.25 and 0.35, respectively. This dataset was used to implement transfer learning with both the solid fractions and base material being different from the initial model training data. The distribution of MD-computed elastic constants and the target solid volume fractions are shown in Figure \ref{computed_modulus} and \ref{solid_fraction}, respectively.

\subsection{Training details}

\subsubsection{$k$-fold cross validation}

Given the relatively small dataset, an $k$-fold cross-validation technique was employed, with $k=8$.
The dataset was randomly partitioned into $k$ non-overlapping subsets, with each subset serving as the test set once, once as a validation set, while the remaining $k-2$ subsets were used for training. In each iteration, a specific fold (e.g., fold 1) was designated as the test set, one specific fold (say fold 2) was used as a validation set, while the remaining six folds were used for training. During the training, in each epoch, the performance of the model was evaluated on the validation set, and the training continued as long as the metrics on the validation set improved. Next, the prediction of the Young modulus for the structures in the test set is made, and the final performance of the model is computed. This process was repeated for all $k$-folds, ensuring that each system was used exactly once as a test sample, once as a validation set, and $k-2$ times as part of the training set. Using cross-validation also enables the computation of the standard deviation across different runs, providing insight into the model's stability and variability.

\subsubsection{Structure stratification}\label{stratification}

In the dataset, each structure was independently compressed along three orthogonal directions. Consequently, for each nanoporous configuration, the apparent Young's modulus was calculated along the three principal axes. This procedure inherently introduced correlations between the computed Young's moduli in different directions. If the data were subjected to a random train/test split, it would lead to data leakage, compromising the integrity of the model evaluation.
To mitigate this issue, a rigorous stratification policy was implemented during training, wherein the splits in the $k$-fold cross-validation were performed based on the structures rather than the computed Young's modulus values. This ensured that all three Young's modulus values derived from a given structure (corresponding to the x-, y-, and z-axes) consistently appeared in either the training, validation, or test set. Consequently, scenarios where, for example, the Young's modulus along the x-axis was included in the test set while the modulus along the y-axis remained in the training set were strictly avoided.
This stratification strategy effectively eliminates data contamination, ensuring a robust separation of the dataset. As a result, the model is compelled to capture the underlying relationship between the structural configuration and its elastic properties without being influenced by inter-axis correlations within individual structures.

\subsubsection{Data augmentation}
To enhance structural diversity and ensure translational robustness of the predictions, random roll augmentation \cite{roll_augmentation} was applied during training. Specifically, each structure underwent circular shifting of voxels along the two dimensions perpendicular to the stress direction. The proportion of shifted voxels was randomly sampled from a uniform distribution, with a maximum shift of 30\%. This augmentation was exclusively applied to the training set, promoting translational invariance while maintaining deterministic predictions.

\subsection{3D-CNN architectures performance compared with the baseline model}

Table \ref{tab:overall_metrics} summarizes the performance metrics of the convolutional architectures together with the baseline fully-connected neural net. The regression metrics reported are: coefficient of determination, mean absolute error, and root mean squared error, all averaged over 8 cross-validation folds.

\begin{table}[h]
\caption{Performance metrics for different CNN architectures considered. The $R^2$, MAE, and RMSE metrics were computed for predictions made on all folds. The best result is denoted with bold font.
The standard deviations, denoted by $\sigma^2$, were computed for metrics obtained in each fold separately.}
\label{tab:overall_metrics}%
\begin{tabular}{l|rrrrrr}
\toprule
CNN Architecture & $R^2$ & $\sigma^2(R^2)$ & MAE & $\sigma^2($MAE$)$ & RMSE & $\sigma^2($RMSE$)$ \\ \hline
Fully-Connected NN & 0.7039 & 0.0135 & 0.2526 & 0.0063 & 0.3387 & 0.0104 \\
MobileNet & 0.9209 & 0.0251 & 0.1317 & 0.0165 & 0.1752 & 0.0217 \\
ResNet-18 & 0.9127 & 0.0156 & 0.1394 & 0.0136 & 0.1840 & 0.0168 \\
ResNet-50 & 0.8947 & 0.0238 & 0.1543 & 0.0174 & 0.2022 & 0.0229 \\
ResNet-101 & 0.8928 & 0.0194 & 0.1550 & 0.0095 & 0.2040 & 0.0132 \\
DenseNet-121 & 0.9491 & 0.0046 & 0.1069 & 0.0051 & 0.1406 & 0.0062 \\
DenseNet-169 & 0.9499 & 0.0130 & 0.1057 & 0.0121 & 0.1394 & 0.0155 \\
DenseNet-201 & \textbf{0.9546} & 0.0065 & \textbf{0.1003} & 0.0084 & \textbf{0.1328} & 0.0110 \\
DenseNet-264 & 0.9424 & 0.0113 & 0.1132 & 0.0107 & 0.1495 & 0.0137 \\
\bottomrule
\end{tabular}
\end{table}

Fully-connected network trained on a set of precomputed descriptors demonstrated the lowest metrics across all the models. While being a baseline solution, it is substantially outperformed by architectures with convolutions. The key limitation of the FCNN model is the expressive capability of the descriptors, which are essentially summarized and averaged statistical information about the topology. Therefore, important local or global patterns contributing to overall stiffness could be undetected and not reflected in a descriptor value.

MobileNet's metrics proved that the lightweight architecture is comparable to much deeper networks in terms of predictive capabilities. As it has a much lower number of parameters to train, MobileNet is less likely to overfit, which is important when training on a small dataset. ResNet-18 showed similar results while having lower metric deviations between folds, suggesting it is less sensitive to training data splitting. The ResNets in general illustrate the problem of diminishing returns - ResNet-50 and ResNet-101 add more layers and parameters to train without improving the accuracy. This indicates that additional training parameters do not always contribute to more robust performance but can lead to overfitting.

The class of DenseNets showcased the efficiency of the dense connections strategy. By combining feature maps of all the previous layers, DenseNet-201 achieved the best metrics overall ($R^2$ = 0.9546, MAE = 0.1003 GPa, RMSE = 0.1328 GPa) with one of the lowest standard deviations, indicating the generalization ability. In general, DenseNet-121, -169, -201, and -264 achieved nearly identical performance, proving again that after a certain model depth, adding new layers brings no performance gains. 

We provide additional information and analysis regarding the best-performing model (DenseNet-201) in the Supplementary Information. 

\subsection{Model performance analysis}

\subsubsection{Influence of smaller grid resolutions}

In the previous section of this study, a voxelization resolution of $80\times80\times80$ was used. To assess how the results change when using a lower-resolution voxelization, the structures were linearly rescaled to resolutions of $60\times60\times60$, $40\times40\times40$, and $30\times30\times30$.
Using the DenseNet-201 architecture, which yielded the best performance according to Table \ref{tab:overall_metrics}, models were trained on the lower-resolution data. The resulting $R^2$ coefficients were 0.9377 for $60\times60\times60$, 0.9067 for $40\times40\times40$, and 0.8791 for $30\times30\times30$. Notably, the decline in performance metrics is relatively small, considering the drastically lower resolution of the structures. The total number of voxels is nearly 19 times lower when using $30\times30\times30$ voxelization instead of the initial $80\times80\times80$, yet the observed decrease in the $R^2$ coefficient ranges only from 0.9546 to 0.8791.
This minor reduction in method performance arises from the fact that the convolutional networks used do not require the input structures to be strictly binary masks, where a value of 1 indicates material presence and 0 denotes a vacuum. During structure scaling, a bilinear interpolation algorithm was employed, introducing fractional voxel occupancy. These fractional values are interpreted by the model as indicating that the voxel does not entirely represent a homogeneous material fill. Consequently, despite the apparent reduction in data granularity, a substantial portion of the microscopic information is preserved, allowing the model to maintain high predictive accuracy even at lower voxel resolutions.

\subsubsection{Influence of dataset size}
The dataset size used in this study was initially chosen without specific optimization and may be larger than necessary for training a reliable model. To investigate the impact of dataset size on model performance, we systematically evaluated the model across varying training set sizes. Subsets containing between $n'=50$ and $n'=2000$ unique structures were randomly sampled from the full dataset. Each structure contributes three compressed variants (corresponding to orthogonal directions), resulting in a total of 150 to 6000 data points per subset. For each dataset size, 75\% of the data was used for training, 12.5\% for validation, and 12.5\% for testing. Although cross-validation was not employed, five independent random samples were generated for each size to ensure statistical robustness.

\begin{figure}
    \centering
    \begin{subfigure}[b]{0.495\linewidth}
    \includegraphics[width=\linewidth]{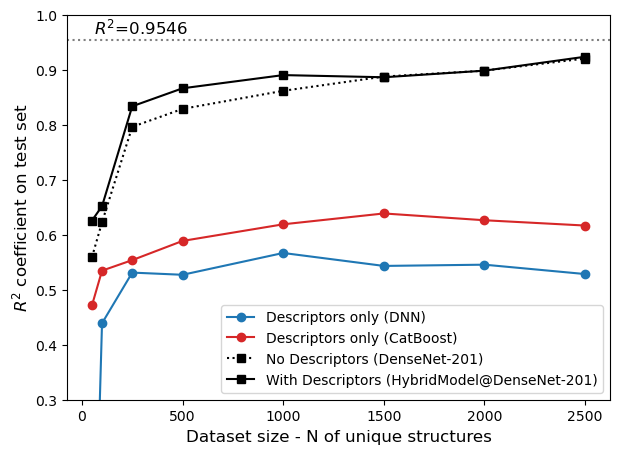}
    \caption{DenseNet-201 based.}
    \end{subfigure}
    \begin{subfigure}[b]{0.495\linewidth}
    \includegraphics[width=\linewidth]{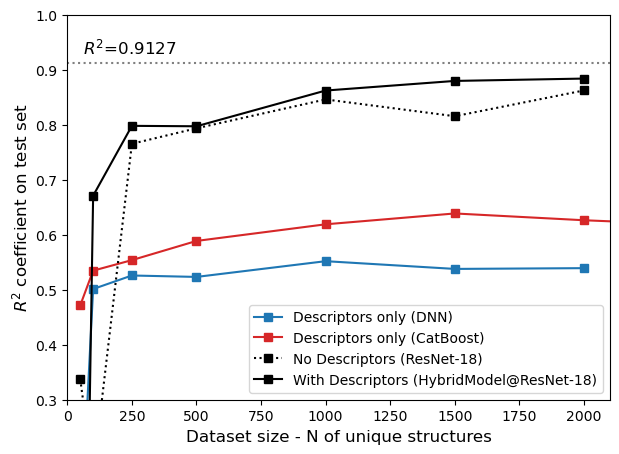}
    \caption{Simpler ResNet-18 based.}
    \end{subfigure}
    \caption{Coefficient of determination ($R^2$) as a function of dataset size.
The black dotted line represents the CNN model without additional descriptors (see Table \ref{tab:overall_metrics}), while the black solid line shows performance with additional descriptors. The blue and red lines show the performance of models based solely on handcrafted descriptors — blue for the fully connected neural network (FC-NN), and red for the gradient boosting algorithm. Each data point represents the average $R^2$ over five different datasets of the given size.}
    \label{fig:hybrid}
\end{figure}

Figure \ref{fig:hybrid} presents the average coefficient of determination ($R^2$) as a function of dataset size $n'$. As shown in Figure \ref{fig:hybrid}a, DenseNet-201 consistently improves performance with increasing dataset size. Figure \ref{fig:hybrid}b confirms that this trend holds when using a different convolutional backbone — ResNet-18 — which, despite delivering lower overall predictive performance than DenseNet-201, shows qualitatively similar behavior across all configurations and dataset sizes. Notably, the DenseNet-201 trained on the dataset reduced by 50$\%$ experienced only a 3.58$\%$ reduction in $R^2$, 9.74$\%$ for 80$\%$, and 16.60$\%$ for 95$\%$ dataset reduction. These results again highlight the data efficiency and generalization abilities of 3D CNNs in general and DenseNets in particular.

\subsubsection{Influence of combining descriptors and grids}
In order to combine the structural information with the descriptors, a custom neural network architecture was proposed. The model integrates information from two complementary sources to make a final prediction: structured data, processed as before by the convolutional neural network, and a list of features that are processed by a fully connected neural network with three hidden layers. The numerical features produced by those two branches are then concatenated and passed through another fully connected neural network to produce the final prediction. The whole architecture is trained end-to-end with weights of all the involved networks being updated during the training. Please refer to Section 3 in the Supplementary Information for a more detailed description and architecture diagram. Overall, this hybrid architecture is designed to jointly learn from both structured (i.e. 3D voxelized structures) and tabular data (i.e. human-engineered features), allowing the model to leverage complementary information from both domains and improve the performance on the regression task.

Figure \ref{fig:hybrid} (red lines) includes the performance of another model variant that relies exclusively on the descriptor processing branch (see Section 3 in the Supplementary Information), omitting any structural input. This descriptor-only model performs worse than either of the CNN-based architectures and exhibits early saturation, with limited gains as the dataset size increases. Additionally, a gradient boosting model using the CatBoost algorithm \cite{prokhorenkova2018catboost} was trained solely on descriptor data (see blue line in Figure \ref{fig:hybrid}). As expected, the boosting model outperforms the neural network trained on descriptors alone—an outcome consistent with existing literature, where tree-based methods often provide strong baselines for tabular data \cite{grinsztajn2022tree, shwartz2022tabular}. In this context, the performance of the gradient boosting model can be viewed as an empirical upper bound for models relying exclusively on descriptors.

Finally, comparing the DenseNet-201 augmented with descriptor information (hybrid model) to the standard DenseNet-201 implementation revealed that their performance is practically the same across all sizes, with some relative improvement from descriptor augmentation observed at smaller dataset sizes. This suggests that descriptor information is particularly beneficial in low-data regimes, where structural patterns alone may be insufficient for optimal elastic constants prediction. A similar behavior was observed for the ResNet-18 hybrid model. The only difference was at the extremely small datasets (50 and 100 structures), where the CatBoost algorithm trained on the descriptors outperformed ResNet-18.

\subsection{Transfer learning}

Among all of the architectures tested in this work, the DenseNet-121 was selected as the base model for transfer learning. While demonstrating nearly identical metrics of predictions compared to the best DenseNet-201 model (Table \ref{tab:overall_metrics}), it has fewer parameters to train, and therefore lower computational costs.

At first, the performance of DenseNet-121 trained on Dataset 1 was assessed on Dataset 2. Metrics of prediction on Dataset 2 are: $R^2$ = 0.712, MSE = 0.593 GPa,  MAE = 0.621 GPa. As it was expected, the model trained on a dataset with a lower solid fraction of 0.25 underestimates the mechanical property for the new set of thicker structures. This is a typical behavior of neural networks and machine learning models in general; the models work best for interpolation and not extrapolation. The difference between MD-computed and predicted values is greater for structures with a higher solid fraction, as their properties are further away from the data the model was trained on.

Next, instead of training a new DenseNet-121 with randomly initialized weights, we employed a transfer learning approach. According to the metrics on Dataset 1, the model already gained knowledge of the structure-property relationship in NPG samples, and it only required fine-tuning of the weights to produce meaningful predictions on Dataset 2. To achieve this, we ran the training again using the data from Dataset 2 but loading the model state from the previously trained DenseNet-121. The dataset was separated into 5 cross-validation folds, mean metrics on the test sets are: $R^2$ = 0.962, MAE = 0.220 GPa, RMSE = 0.2771 GPa with $\sigma^2(R^2)$ = 0.0079, $\sigma^2(MAE)$ = 0.0213 and $\sigma^2(RMSE)$ = 0.0279. The metrics show that transfer learning can effectively solve the problem of limited training data. With only 1000 structures available from Dataset 2, the model achieved the performance of the original training.

\begin{figure}[h]
    \centering
    \includegraphics [width=0.75\textwidth]{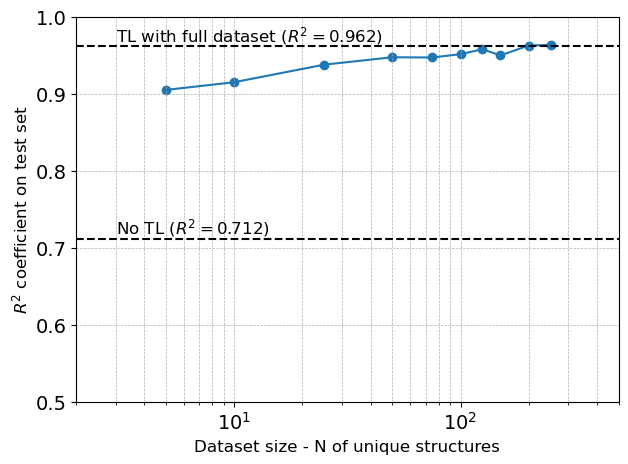}
    \caption{$R^2$ test score as a function of fine-tuning structures number.} 
  \label{df_frac}
\end{figure}

To explore the limit of transfer learning with extremely limited training data, we progressively reduced the fine-tuning dataset size, observing changes in model accuracy. First, 200 structures (600 datapoints) were held out as the test set to evaluate all the models on the same data. From the remaining 800 structures of Dataset 2, we selected subsets of size 5, 10, 25, 50, 75, 100, 125, 150, 200, and 250 structures to fine-tune DenseNet-121. The same stratification strategy described in Section \ref{stratification} was applied to prevent data leakage and ensure, for example, that all 15 datapoints belong to 5 unique structures selected for training. Figure \ref{df_frac} shows the $R^2$ test score as a function of the number of unique structures selected for fine-tuning. Remarkably, even with only 5 unique structures to train on (15 datapoints), the model achieved $R^2$ score of 0.905 on the 200 structures test set. It demonstrates that the initial DenseNet-121 learned the fundamental relationships between ligament network topology and stiffness, only requiring several new examples to adapt the outputs. 

Finally, the pretrained DenseNet-121 was fine-tuned on nanoporous silver structures with two target values of solid fraction from Dataset 3. The data was separated into 5 folds for cross-validation, and the mean test metrics achieved: $R^2$ = 0.9846, MAE = 0.2336 GPa, RMSE = 0.3206 GPa with $\sigma^2(R^2)$ = 0.0033, $\sigma^2(MAE)$ = 0.0259 and $\sigma^2(RMSE)$ = 0.0422. These metrics confirm the robustness of the learned features and demonstrate that transfer learning can be applied to a broad range of nanoporous metal systems.

\subsection{Generalization test on topologically defined structures}

To test the model's performance beyond the training dataset, we evaluated the DenseNet, trained on stochastic structures only, on a separate set of architected topologies collected from the RCSR database (see \cite{zorkaltsev2025molecular} for the topologies description and the structures generation procedure). This new set of structures included 22 topologies, each generated with 5 different values of solid fraction, which resulted in a total of 110 structures labeled with MD-computed values of elastic constant.
As it was expected, absolute prediction errors on these defined structures are substantially larger than on the stochastic test set, because the network was never trained on such ordered samples. However, the model was able to maintain the ranking of stiffness across these different topologies. The Spearman rank coefficient between MD results and predicted stiffness is $\rho$ = 0.955 with a p-value of 3.26$\times10^{-58}$). This strong agreement indicates that the structure–property patterns learned from stochastic training data are useful for topologically defined structures, although quantitative predictions on architected samples will require further tuning with transfer learning. Detailed results and example structures are provided in the Supplementary Information.

\subsection{Screening of random NPG structures}

\begin{figure}[h]
    \centering
    \includegraphics [width=0.8\textwidth]{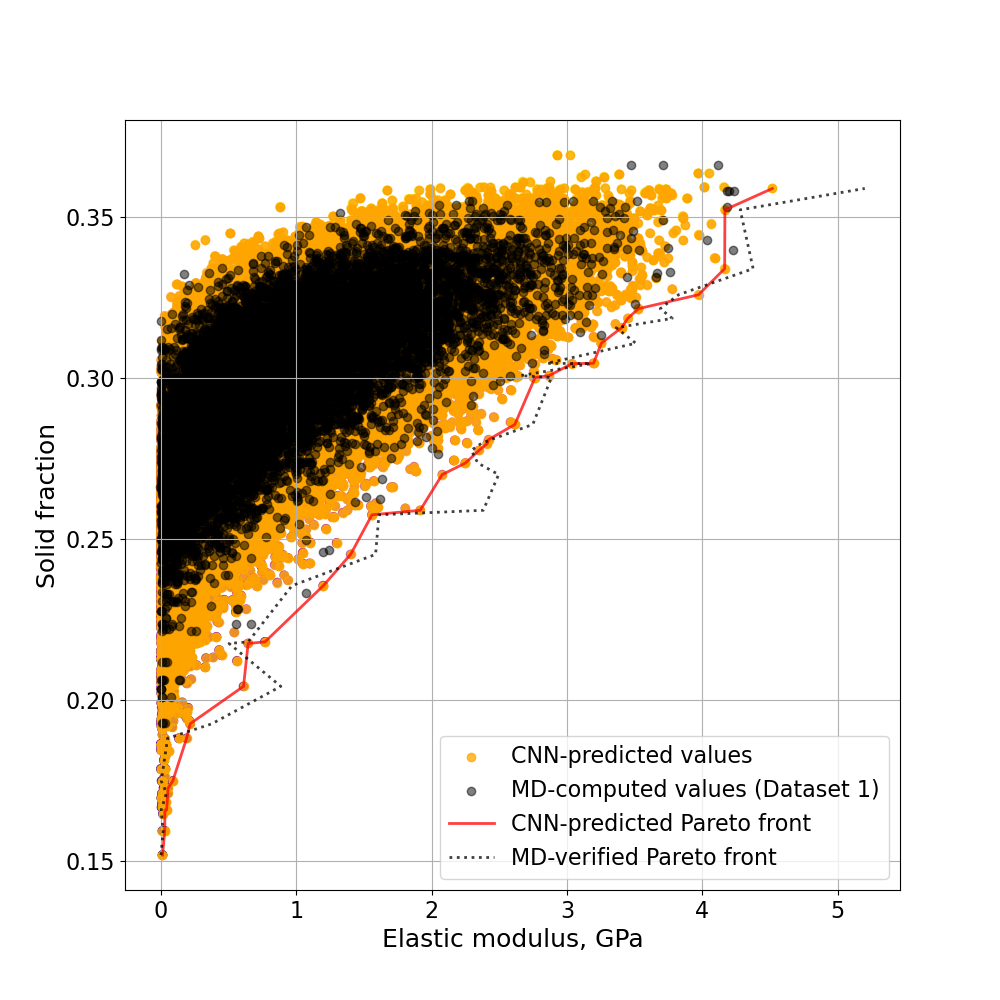}
    \caption{CNN-predicted elastic modulus values of 100.000 NPG structures plotted against solid fraction, with MD-computed values from Dataset 1 as a reference.} 
  \label{pareto}
\end{figure}

To explore the design space of NPG structures and possibly identify the designs with higher elastic modulus, the pretrained DenseNet-121 was applied to evaluate 100,000 randomly generated structures in the three principal directions. The resulting 300.000 CNN-predicted elastic modulus values were plotted against solid fraction (Figure \ref{pareto}). As a reference, the plot includes MD-computed elastic modulus values from Dataset 1 used for model training. 

 Pareto front optimal structures were extracted from the scatter, defined by the highest property value for a given solid fraction. Additional MD simulations were run to verify CNN predictions for 33 structures forming a Pareto front. The simulations confirmed strong agreement with CNN predictions: $R^2$=0.9787, RMSE = 0.2216 GPa, MAE = 0.1659 GPa. The highest deviation between predicted and calculated elastic modulus values was observed for the top-performing structure (predicted 4.52 GPa and MD-verified 5.20 GPa). This is expected behavior of the model, as the top structure lies on the boundary of the training data distribution for both elastic constant and solid fraction, making extrapolation harder for the model.

\section{Conclusion}\label{Conclusions}

In this study, we have demonstrated that 3D Convolutional networks offer a remarkable advantage over traditional descriptor‐based models by directly learning from local and global topological features of nanoporous metals. The generated datasets included 6,422 structures in total, each labeled with MD-computed elastic constants in three principal directions. Among the tested architectures, DenseNets in general and DenseNet‑201 in particular achieved outstanding accuracy of predictions and metrics stability with 8 cross-validation folds.

Several key factors affecting the model performance were evaluated through a series of experiments. First, it was demonstrated that the DenseNet is remarkably robust to an input grid resolution. Even reducing the total number of voxels 19 times (from $80^3$ to $30^3$ voxels resolution), the determination coefficient $R^2$ decreased by only ~7.5$\%$. Second, the analysis of varying dataset sizes revealed that the models maintain their performance when trained on a fraction of the original dataset. For instance, training the DenseNet-201 on 50$\%$ of the original structures led to less than 4$\%$ drop in $R^2$. Additionally, a hybrid approach combining the grid representation with topological and morphological descriptors was proposed and evaluated. Although the hybrid models showed minimal gain compared to the standard architectures in larger datasets, they demonstrated improvements in the smaller training sets. 

The investigation of transferability further revealed the robustness of representations learned by the models. A DenseNet trained on 5,000 gold samples was effectively fine‑tuned with several new data points to achieve reliable predictions on the denser structures. Moreover, a model trained on one material can be easily adapted to a different one (e.g., silver) with minimal data required. This data efficiency is important in materials research or other fields where generating large labeled datasets involves computationally expensive simulations. By employing our trained DenseNet as a screening tool, we evaluated 100,000 stochastic NPG structures and identified the Pareto optimal designs, which were validated with additional MD simulations. 

We highlight several promising directions in which this work can be extended. The first one is to calibrate the models trained on the MD-labeled dataset to other reference data (e.g., DFT calculations, FE simulations, or experiments), which was outside the scope of this study. It would further demonstrate the transferability of learned patterns across different training data labeling techniques. Second, although we demonstrated generalization on 110 architected samples from the RCSR database, this topic could be further extended to achieve the same metrics as on stochastic samples demonstrated in this study. Moreover, training the CNN architecture that shows stable high accuracy on both stochastic and ordered samples of, possibly, different base materials could reduce the need to train and fine-tune a new model for every specific case. Third, we emphasize that the trained DenseNet acts as a fast surrogate model and rapid screening tool for stochastic structures, substantially reducing the time for elastic constant calculation compared to MD simulations. SHAP analysis (Supplementary Information, Section 3.1) showed it is possible to identify regions contributing to higher/lower stiffness values. This justifies using the CNN as a guide for targeted modifications, but systematic topology optimization requires integrating network suggestions with additional MD evaluations and is therefore left as a subject for future work.

In summary, our work presents a robust and tunable pipeline for elastic constant prediction and optimization in porous materials. We demonstrated that Deep learning models are applicable in various scenarios, including limited available data, coarse voxel resolution, and different training data representations. 

\section*{Declarations}

\begin{itemize}
\item \textbf{Author contributions:} 
S.Z: methodology, software (3D-CNNs), validation, investigation, formal analysis, data curation, writing - original draft; 
R.T.: methodology, software (3D-CNNs), validation, investigation, formal analysis, data curation, writing – review and editing;
T.C.: software (structures generation, MD simulations, descriptors calculation), validation, investigation, formal analysis, data curation; 
S.M.: software (structures generation, MD simulations, descriptors calculation), validation, investigation, formal analysis; 
P.D.: conceptualization, supervision; 
D.M.: conceptualization, supervision; 
M.H.: methodology, conceptualization, supervision, project administration, funding acquisition.
All authors: writing — review and editing.
\item \textbf{Acknowledgments:} 
Authors acknowledge the financial support from M-ERA.NET’s PORMETALOMICS project supported by MCIN/AEI/10.13039/501100011033 and the European Union’s NextGenerationEU/PRTR funds. Financial support from the PORMETALOMICS project, funded by the National Science Centre, Poland (project no. 2021/03/Y/ST5/00232) within the M-ERA.NET 3 programme, is also gratefully acknowledged. This project has received funding from the European Union’s Horizon 2020 research and innovation programme under grant agreement No 958174.

The computations in this work were partially supported by the Center for Artificial Intelligence at Adam Mickiewicz University. We would like to thank Prof. K. Jassem and Dr. B. Naskrecki for granting us access to the Center’s infrastructure.

\item \textbf{Competing interests:} All authors declare no financial or non-financial competing interests.
\item \textbf{Data and Code availability:} The data and code are available in a public repository on \href{https://github.com/RafalTopolnicki/cnn-nanoporous}{GitHub}.
\end{itemize}

\clearpage
\clearpage
\setcounter{section}{0}
\setcounter{figure}{0}
\setcounter{table}{0}
\setcounter{equation}{0}
\setcounter{algorithm}{0}

\renewcommand{\thesection}{S\arabic{section}}
\renewcommand{\thefigure}{S\arabic{figure}}
\renewcommand{\thetable}{S\arabic{table}}
\renewcommand{\theequation}{S\arabic{equation}}
\renewcommand{\thealgorithm}{S\arabic{algorithm}}

\begin{center}
{\Large\bfseries Supplementary Information\par}
\end{center}

\vspace{5em}

\section{Generation of nanoporous structures}
\label{generate_NPG}
\renewcommand{\theequation}{S\arabic{equation}}
\renewcommand{\thefigure}{S\arabic{figure}}

Atomistic 3D nanoporous configurations were generated by implementing the algorithm proposed by Soyarslan et al. \cite{SOYARSLAN}, which involves superposition of standing sinusoidal waves of fixed wavelength, random phase, and specific directions. The corresponding random field generator is mentioned in Eq. \ref{eq:random_gen}

\begin{equation}
    \begin{aligned}
    f(\vec{x}) = \sqrt{\frac{2}{N}} \sum_{i=1}^{N} \cos\left( \vec{q}_i \cdot \vec{x} + \phi_i \right)
    \end{aligned}
    \label{eq:random_gen}
\end{equation}

where $\vec{x}$ is the position vector, $\vec{q}_i$ and $\phi_i$ denote the wave direction and phase, respectively and $N$ indicates the number of waves involved in the generation of the nanoporous microstructure. A specific set of vectors $\vec{q}_i=\frac{2\pi}{a}(h,k,l)$ was defined based on a constant value $H = \sqrt{h^2 + k^2 + l^2}$, where $h$, $k$, $l$ are integers corresponding to the Miller indices and $a$ is the edge length of the simulation domain, and therefore $|\vec{q}| = q_0 = \frac{2\pi H}{a}$. 

In this study, $H^2 = 9$, the number of repetitions of the unit cell = 125, and the domain size is given by $a = \text{lattice constant} \times 125$. Accordingly, the 3D periodic nanoporous structures were generated using the random field defined in Eq. \ref{eq:random_gen}. For $H^2 = 9$, there are $30$ wave directions and they are mentioned in Table \ref{tab:vector_list}.

\begin{table}[h]
\renewcommand{\thetable}{S\arabic{table}}
\caption{Wave direction vector list corresponding to $H^2 = 9$}
\label{tab:vector_list}
\begin{tabular}{|c|c|c|c|}
 \hline
Index & h & k & l \\
\hline
1  & -3 & 0  & 0  \\
2  & -2 & -2 & -1 \\
3  & -2 & -2 & 1  \\
4  & -2 & -1 & -2 \\
5  & -2 & -1 & 2  \\
6  & -2 & 1  & -2 \\
7  & -2 & 1  & 2  \\
8  & -2 & 2  & -1 \\
9  & -2 & 2  & 1  \\
10 & -1 & -2 & -2 \\
11 & -1 & -2 & 2  \\
12 & -1 & 2  & -2 \\
13 & -1 & 2  & 2  \\
14 & 0  & -3 & 0  \\
15 & 0  & 0  & -3 \\
16 & 0  & 0  & 3  \\
17 & 0  & 3  & 0  \\
18 & 1  & -2 & -2 \\
19 & 1  & -2 & 2  \\
20 & 1  & 2  & -2 \\
21 & 1  & 2  & 2  \\
22 & 2  & -2 & -1 \\
23 & 2  & -2 & 1  \\
24 & 2  & -1 & -2 \\
25 & 2  & -1 & 2  \\
26 & 2  & 1  & -2 \\
27 & 2  & 1  & 2  \\
28 & 2  & 2  & -1 \\
29 & 2  & 2  & 1  \\
30 & 3  & 0  & 0  \\
\hline
\end{tabular}
\end{table}

\section{Extraction of morphological and topological descriptors}
\label{si:topo_morpho_descrip}

\begin{figure}[h]
    \centering
    \begin{subfigure}[b]{0.4\textwidth}
        \includegraphics[width=\textwidth]{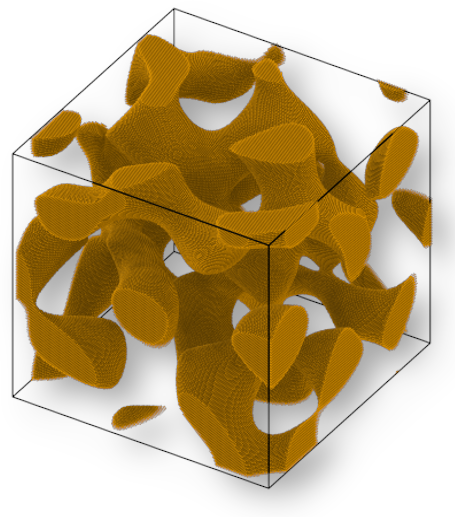}
        \caption{}
        \label{fig:NPGstructure}
    \end{subfigure}
    \hfill
    \begin{subfigure}[b]{0.4\textwidth}
        \includegraphics[width=\textwidth]{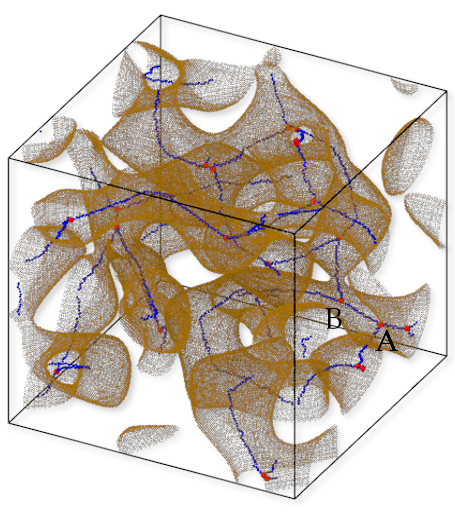}
        \caption{}
        \label{fig:skeleton}
    \end{subfigure} \newline
    \begin{subfigure}[b]{0.4\textwidth}
        \includegraphics[width=\textwidth]{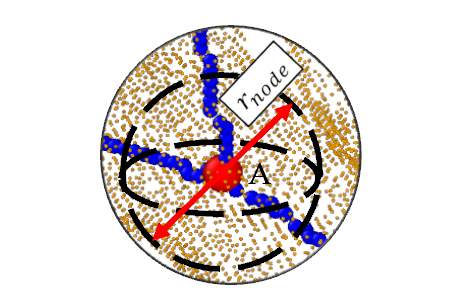}
        \caption{}
        \label{fig:node}
    \end{subfigure}
    \hfill
    \begin{subfigure}[b]{0.4\textwidth}
        \includegraphics[width=\textwidth]{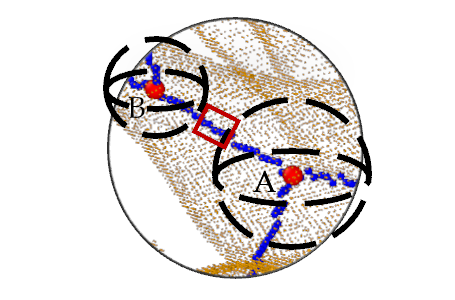}
        \caption{}
        \label{fig:node_ligament}
    \end{subfigure}
    \caption{Topological descriptors. (a) An atomic nanoporous structure. (b) The skeleton, superimposed with the surface atoms of the nanoporous structure. (c) The sphere defines the size of the node at point A (marked in (b)). (d) Two adjacent nodes (A and B, marked in (b)) and the skeleton elements are considered as the ligaments in between the nodes.}
    \label{fig:descriptors}
\end{figure}

The morphological and topological descriptors of the nanoporous structure were extracted from the skeletonized network, which was obtained through post-processing of the initially generated 3D nanoporous models. First, a 3D image of the nanoporous structure was created using voxels, with each voxel sized approximately equal to the diameter of an atom. The voids between the voxels were filled to ensure a continuous and well-defined representation of the porous network. A 3D skeletonization algorithm \cite{Lee94_skeleton} was then applied to this voxel-based model, reducing it into a skeleton framework while preserving the topology of the porous network. Finally, the skeletonized structure was converted back into coordinate data, enabling further analysis and visualization. The descriptors were then derived by integrating the skeleton elements and nodes with the corresponding nanoporous volume. An example of a fully periodic nanoporous structure from atomistic simulations is presented in Fig. \ref{fig:descriptors}a. The corresponding skeleton, segmented into discrete points, is shown in \ref{fig:descriptors}b, where points located at node centers are highlighted in red.

Leveraging the skeleton, several directional descriptors are evaluated in this study. The first among them are the ligament-only directional resistance (LDR) and its modified form (MLDR), both defined in the main text, similar to one defined in Ref. \cite{Mathesan}. These descriptors assume that all skeleton elements, connecting pairs of skeleton points, constitute ligaments, thereby neglecting the finite size of the nodes. The descriptors are defined as

\begin{equation}
    \begin{aligned}
        LDR &=& \frac{\sum_{ligaments} l_i \cos{\gamma_i}}{L}, \\
        MLDR &=&\frac{\sum_{ligaments} l_i \cos^{10}{\gamma_i}}{L},
    \end{aligned}
    \label{eq:LDR_MLDR}
\end{equation}
where $l_i$ is the length of the skeleton elements, $\gamma_i$ is the angle between the skeleton element and the loading axis, and $L$ is the size of simulation box along the loading axis.

The previous approach implicitly assumes that the mechanical response of skeleton elements is uniform across ligament centers and node centers, while nodes contributing an isotropic response to deformation, as opposed to ligaments. To account for this, we reclassified skeleton elements, distinguishing those corresponding to ligaments from those residing within nodes. Specifically, a node is defined as the largest sphere centered on a skeleton node and fully contained within the nanoporous volume. An illustrative example is provided in Fig. \ref{fig:descriptors}c, for the node labeled A in Fig. \ref{fig:descriptors}b. Skeleton points lying outside these spheres are identified as ligaments, such as those within the red rectangle in Fig. \ref{fig:descriptors}d, corresponding to the ligament connecting nodes A and B. Notably, in the current study, under this definition, ligaments are substantially shorter than in the previous approach, and their lengths are less than the node size $r_{\text{node}}$. Based on this redefinition, and recognizing that nodes respond isotropically to mechanical loading, we introduce two new directional descriptors: the directional resistance (DR) and the modified directional resistance (MDR),
\begin{equation}
    \begin{aligned}
        DR &=& \frac{\sum_{ligaments} l_i \cos{\gamma_i}+\sum_{nodes}2r_{node,j}}{L},  \\  
        MDR &=& \frac{\sum_{ligaments} l_i \cos^{10}{\gamma_i}+\sum_{nodes}2r_{node,j}}{L}.
    \end{aligned} 
    \label{eq:DR_MDR}
\end{equation}
The summation over ligaments includes only those skeleton elements that lie outside the node volumes, while the summation over nodes includes the full set of skeleton nodes.

In the above definition, ligament contributions are considered independent of their cross-sectional area, implying that thick and thin ligaments contribute equally if oriented at the same angle relative to the loading axis. To incorporate geometric variability, we calculate the mean cross-sectional radius $r_i$, defined perpendicular to each skeleton element classified as a ligament. To capture both length and cross-sectional area, we introduce modified descriptors analogous to Eqs. \ref{eq:LDR_MLDR} and \ref{eq:DR_MDR}, replacing length-based terms with volumetric ones. Accordingly, two new descriptors are defined: ligament-only volumetric resistance (LVR) and modified ligament-only volumetric resistance (MLVR), both excluding contributions from node regions, 
\begin{equation}
    \begin{aligned}
        LVR &=& \frac{\sum_{ligaments} \pi l_i r_i^2 \cos{\gamma_i}}{V}, \\
        MLVR &=&\frac{\sum_{ligaments} \pi l_i r_i^2 \cos^{10}{\gamma_i}}{V}.
    \end{aligned}
    \label{eq:LVR_MLVR}
\end{equation}
Here, $V$ denotes the total volume of the simulation box. Similarly, when the volumes of the nodes are taken into account, we define two additional descriptors: volumetric resistance (VR) and modified volumetric resistance (MVR). These descriptors incorporate both ligament and node contributions, providing a more comprehensive measure of directional mechanical resistance that accounts for topology and the cross-section are of the ligaments.
\begin{equation}
    \begin{aligned}
        DR &=& \frac{\sum_{ligaments} \pi l_i r_i^2 \cos{\gamma_i}+\sum_{nodes}\frac{4}{3}\pi r_{node,j}^3}{V},  \\  
        MDR &=& \frac{\sum_{ligaments} \pi l_i r_i^2 \cos^{10}{\gamma_i}+\sum_{nodes}\frac{4}{3}\pi r_{node,j}^3}{V}.
    \end{aligned}  
    \label{eq:VR_MVR}
\end{equation}
\vspace{0.5cm}

\section{Best-performing architecture (DenseNet-201)}\label{DenseNet-201}

\begin{figure}
    \centering
    \includegraphics[width=0.75\linewidth]{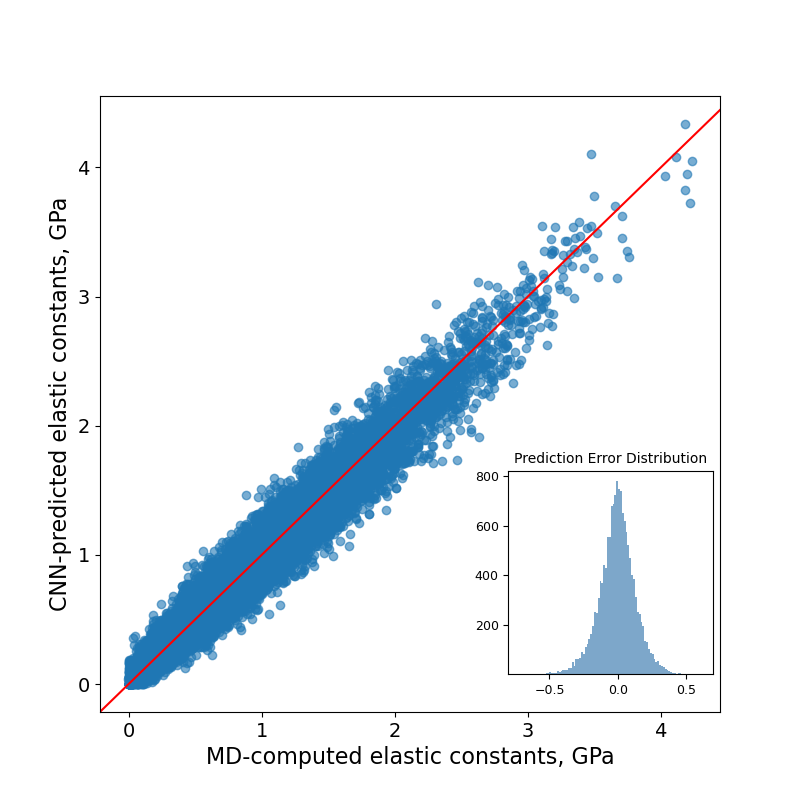}
    \caption{MD-computed vs CNN-predicted elastic constants for 8 cross-validation folds with distribution of errors.}
    \label{fig:scatter}
\end{figure}

The best-performing model in this study is a 3D version of the DenseNet-201 architecture. The network uses an initial convolutional layer with 64 filters followed by four dense blocks with 6, 12, 48, and 32 dense layers, and a growth rate of 32. Dense layers use a bottleneck convolution with a 1$\times$1$\times$1 kernel followed by a 3$\times$3$\times$3 convolution. Transition layers consist of a 1$\times$1$\times$1 convolution and 2$\times$2$\times$2 average pooling. The model maps a single-channel voxelized nanoporous structure to a single scalar output ($C_{ii}$). A scatter plot of MD-computed and predicted values of elastic constants ($C_{11}$, $C_{22}$, and $C_{33}$) is shown in Figure \ref{fig:scatter} together with a histogram of prediction errors.

\begin{figure}
    \centering
    \includegraphics[width=0.85\linewidth]{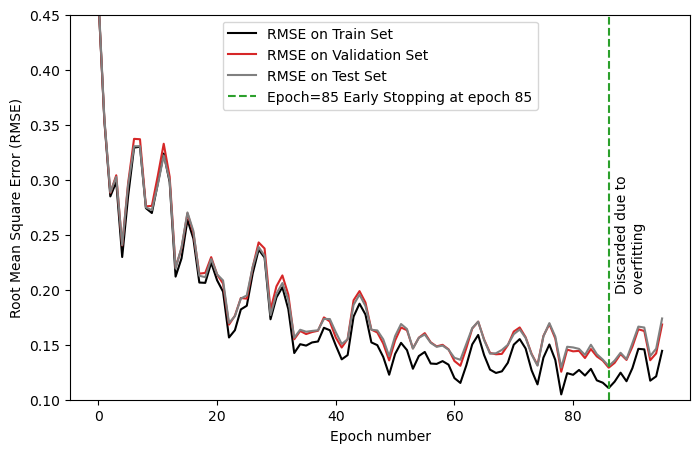}
    \caption{Train, test, and validation loss during the training on Dataset 1.}
    \label{fig:loss}
\end{figure}

Model training was performed with MSE as the loss function. The training loss is computed on the training set, and after each epoch the model is evaluated on the validation set to compute the validation loss. We monitor validation loss to select the best checkpoint: training continues while the validation loss decreases, but it is not stopped at the first validation increase. Instead, we allow a patience of 10 epochs - if the validation loss does not improve below the current best value within that period, training is terminated and the checkpoint corresponding to the lowest observed validation loss is restored. The per-epoch train, test, and validation loss demonstrating stable convergence and the absence of overfitting are shown in Figure \ref{fig:loss}.

DenseNet blocks with dense connectivities promote feature reuse throughout the whole network and ensure gradient flow through deep 3D stacks, enabling training of the model without vanishing gradients. The growth rate determines how many new filters (channels) each layer contributes, while bottleneck convolutions compress intermediate channels to reduce computation and memory costs without sacrificing predictive power. Convolutions with the kernel size of  3×3×3 were used to extract local morphological features. Stacking these small kernels with downsampling increases the receptive field of the model so that deeper layers capture mesoscale and near-global patterns. Transition layers (1$\times$1$\times$1 convolution followed by average pooling) reduce channels and downsample data resolution. Batch normalization is used to stabilize training in the deep 3D network, and a final global average pooling reduces spatial dimensions to a small fully connected regressor that outputs a single value of elastic constant. Optional dropout in the dense layers is used for additional regularization and prevention of overfitting. Layer-by-layer data shapes are provided in Table \ref{tab:layers}.

\begin{table}[h]
\centering
\caption{Layer by layer output shapes and channels of DenseNet-201 for an input size of $80^3$.}
\begin{tabular}{lcc}
\hline
\textbf{Layer} & \textbf{Output shape} & \textbf{Channels} \\
\hline
Input & $80 \times 80 \times 80$ & 1 \\
Conv1 (7$\times$7$\times$7, stride=(1,2,2)) & $80 \times 40 \times 40$ & 64 \\
MaxPool (3$\times$3$\times$3, stride=2) & $40 \times 20 \times 20$ & 64 \\
DenseBlock1 (6 layers) & $40 \times 20 \times 20$ & 256 \\
Transition1 (1$\times$1 conv + AvgPool) & $20 \times 10 \times 10$ & 128 \\
DenseBlock2 (12 layers) & $20 \times 10 \times 10$ & 512 \\
Transition2 & $10 \times 5 \times 5$ & 256 \\
DenseBlock3 (48 layers) & $10 \times 5 \times 5$ & 1792 \\
Transition3 & $5 \times 2 \times 2$ & 896 \\
DenseBlock4 (32 layers) & $5 \times 2 \times 2$ & 1920 \\
BatchNorm + ReLU & $5 \times 2 \times 2$ & 1920 \\
Global AvgPool & $1 \times 1 \times 1$ & 1920 \\
Fully Connected & $1 \times 1$ & 1 (elastic constant) \\
\hline
\end{tabular}
\label{tab:layers}
\end{table}

\subsection{Model interpretability}
\begin{figure}
    \centering
    \includegraphics[width=0.95\linewidth]{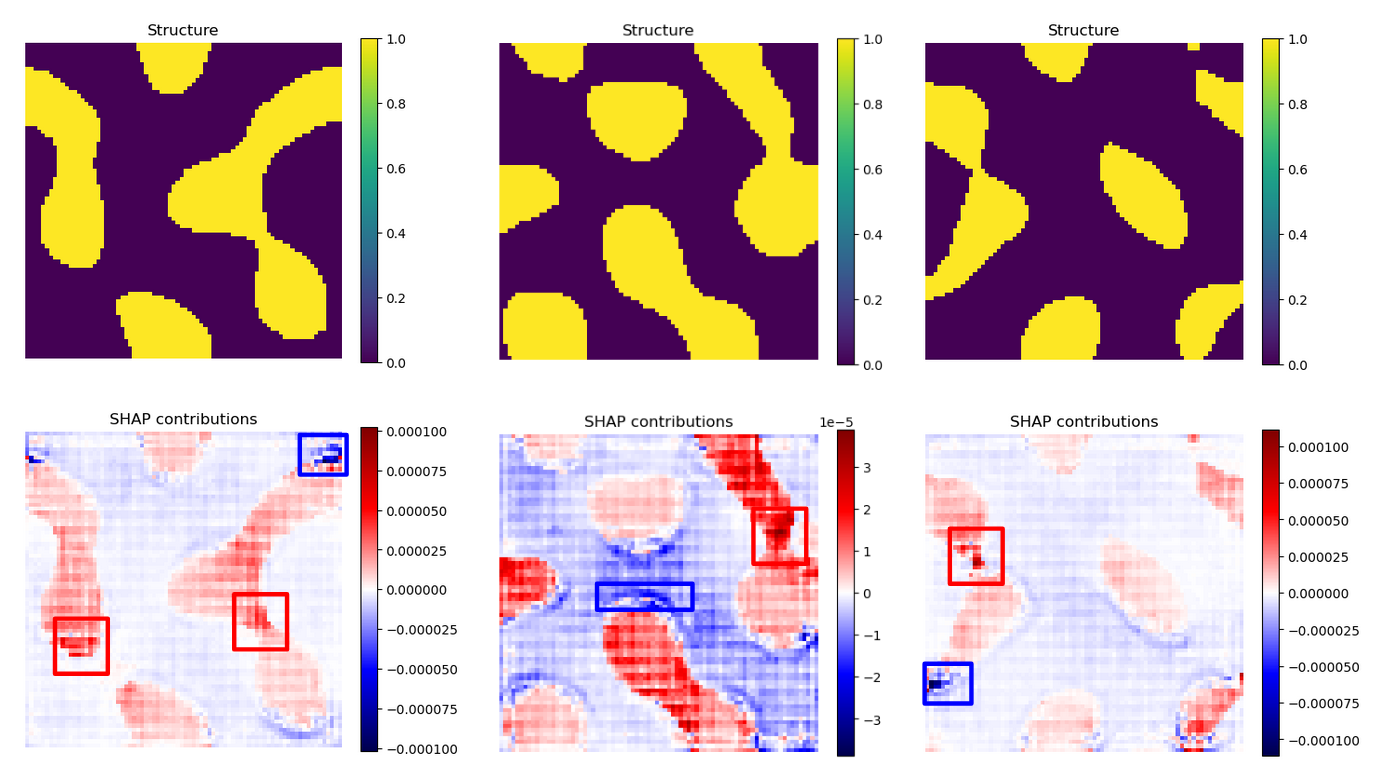}
    \caption{Intersection planes through 3D structure and corresponding spatial distribution of SHAP values. Red (blue) color indicates regions with positive (negative) impact on the predicted sample stiffness.}
    \label{fig:shap}
\end{figure}

In order to understand which parts of the NPG structure have the highest influence on the final prediction, a SHAP analysis of the DenseNet-201 model was performed. SHAP is a feature attribution method based on cooperative game theory, which assigns each input feature a contribution score reflecting its impact on the model’s output \cite{lundberg2017}. For high-dimensional models such as 3D CNNs, exact computation of Shapley values is infeasible, and SHAP employs approximation strategies (e.g., DeepSHAP, \cite{lundberg2017, shrikumar2019}) that propagate attributions through the network layers. In our case, the CNN input is a voxelized representation of the structure, and SHAP values measure the marginal contribution of each structural element to the predicted stiffness. Positive values indicate features that increase the predicted modulus, while negative values correspond to features that decrease it. By aggregating these values across samples, we can identify which structural motifs consistently drive the model’s prediction and thereby possibly gain physical insight.

Figure \ref{fig:shap} illustrates three orthogonal intersection planes through the same structure, taken perpendicular to the direction of the applied stress. In the top row, the binary structure is shown as a 2D cross-section, while in the bottom row the corresponding voxel-resolved SHAP values are plotted. Regions with positive SHAP values, shown in red, denote areas that contribute positively to the overall stiffness of the structure. Several of the most influential regions, highlighted with red boxes, coincide with material-rich areas, particularly narrow struts or thin connections within the network. The high SHAP values in these regions indicate that their removal (i.e., replacement with void) would result in the largest reduction in stiffness, emphasizing their critical role in sustaining the mechanical response of the sample. Put differently, red regions identify the structural motifs with the strongest positive contribution to stiffness.

In contrast, regions with negative SHAP values, shown in blue, are typically located adjacent to pores or vacuum regions. Dark blue zones, highlighted with blue boxes, where the SHAP values are large in magnitude but negative, are most often located near the structure boundaries. Dark blue zones, highlighted with blue boxes, are most often found near the structure boundaries. These negative attributions imply that the presence of a void in these areas has a large negative impact on mechanical strength. The model suggests that adding material here could yield the greatest relative improvement in stiffness. In other words, while these empty voxels currently contribute negatively to the mechanical load-bearing capacity, they represent the regions with the highest potential for improvement: if solid material were introduced here, the structure could achieve a significant enhancement in its overall stiffness.

This interpretation highlights the dual role of SHAP analysis: it not only identifies the structural features most critical for maintaining stiffness but also pinpoints the “weak links” where targeted modifications could most effectively reinforce mechanical performance. Unlike raw molecular dynamics simulations, which primarily provide numerical estimates of properties, the deep learning model, together with SHAP analysis, enables a systematic approach to improving the mechanical strength of a given sample.

\begin{figure}[htbp]
  \centering
  \begin{subfigure}[b]{0.35\linewidth}
    \centering
    \includegraphics[width=\linewidth]{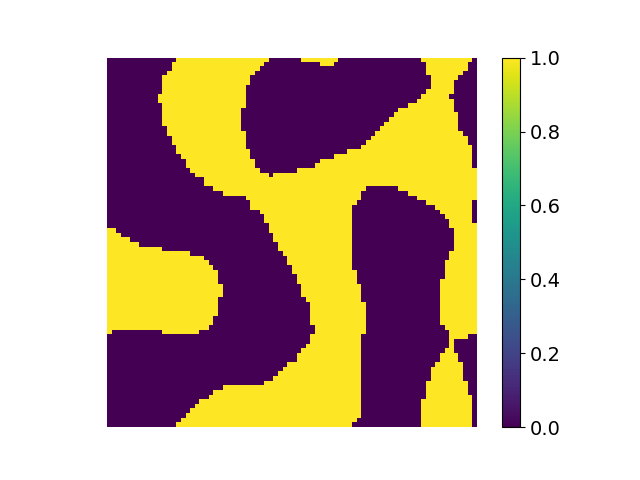}
    \caption{Input tensor slice (binary).}
    \label{fig:layers_a}
  \end{subfigure}
  \hfill
  \begin{subfigure}[b]{0.85\linewidth}
    \centering
    \includegraphics[width=\linewidth]{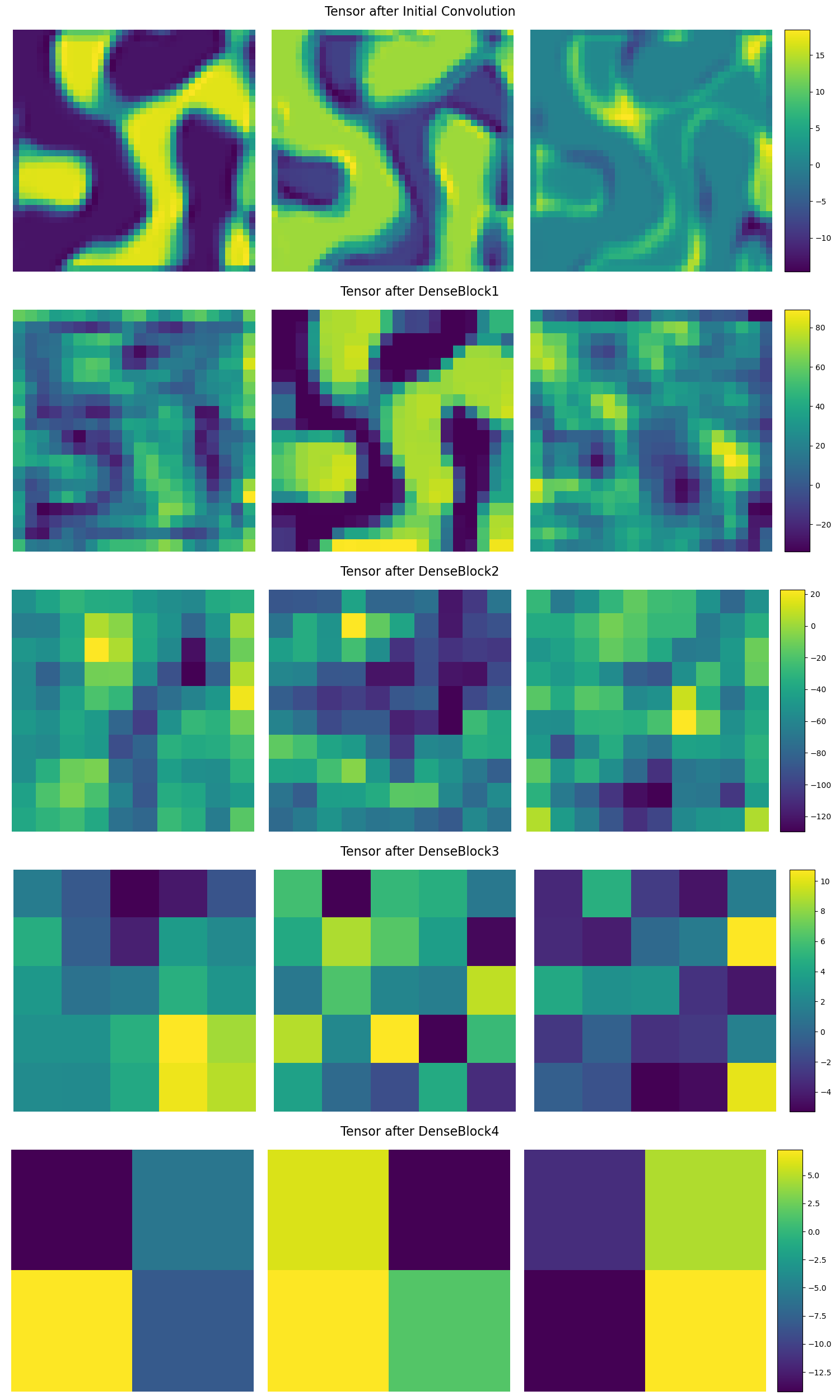}
    \caption{Intermediate-layer activations (Initial Conv + DenseBlocks).}
    \label{fig:layers_b}
  \end{subfigure}
  \caption{Visualization of DenseNet-201 intermediate layers.}
  \label{fig:layers}
\end{figure}

Besides SHAP analysis, we visualized how the input structure's representation transforms inside the model. Figure \ref{fig:layers} shows 2D slices of a voxelized input as it passes through initial convolution and DenseBlocks. The initial convolutional filters (3 out of 64 shown) activate on low-level features: one filter responds primarily to solid regions of a structure, another activates on voids, and a third responds strongly at narrow necks where ligaments connect. After DenseBlock1, some of the activation maps still reflect the initial input geometry, as the network operates on high-resolution data at this shallow depth. Deeper layers of the network become progressively more abstract. The activation volumes observed after DenseBlock2, DenseBlock3, and DenseBlock4 have sizes of 10$\times$10, 5$\times$5, and 2$\times$2, respectively. These low-resolution feature maps do not reflect voxelized nanoporous structure but instead encode higher-level, mesoscale, and global structural features that are relevant for predicting elastic constants. 

\section{Generalization test on architected structures}

\begin{figure}[htbp]
  \centering
  \begin{subfigure}[b]{0.24\linewidth}
    \centering
    \includegraphics[width=\linewidth]{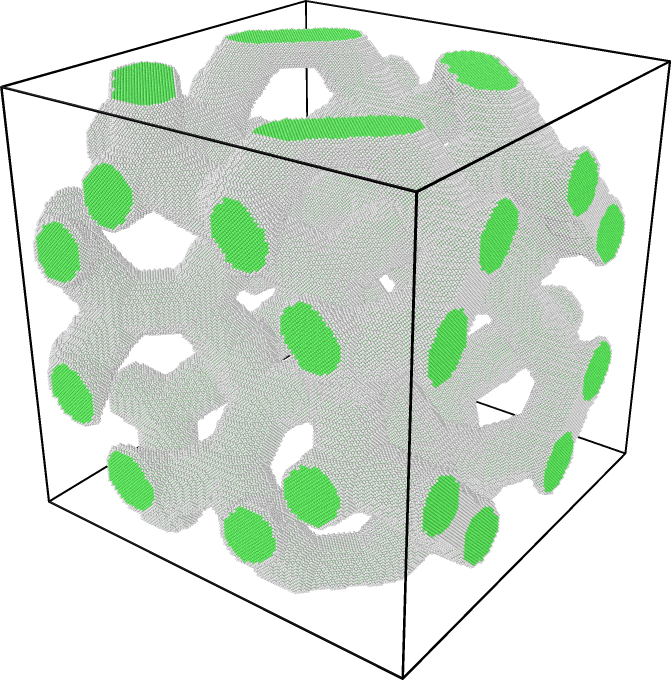}
    \caption{}
    \label{}
  \end{subfigure}
  \hfill
  \begin{subfigure}[b]{0.24\linewidth}
    \centering
    \includegraphics[width=\linewidth]{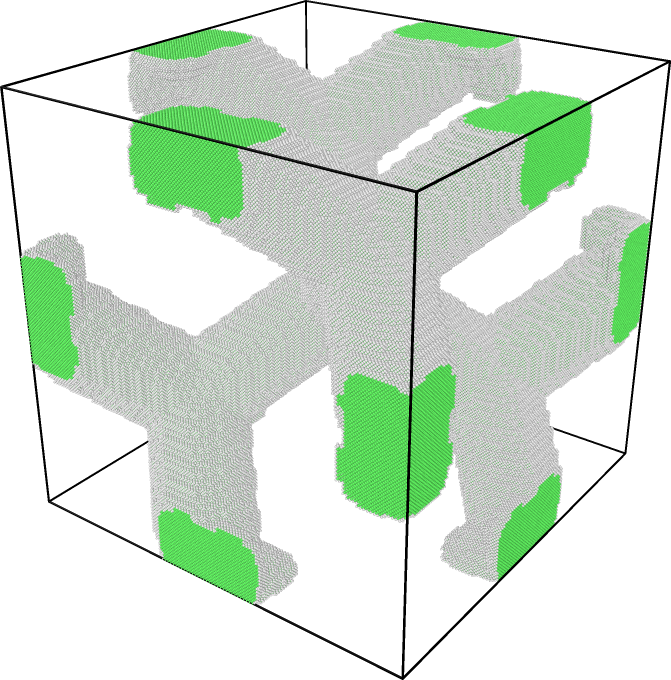}
    \caption{}
    \label{}
  \end{subfigure}
  \hfill
  \begin{subfigure}[b]{0.24\linewidth}
    \centering
    \includegraphics[width=\linewidth]{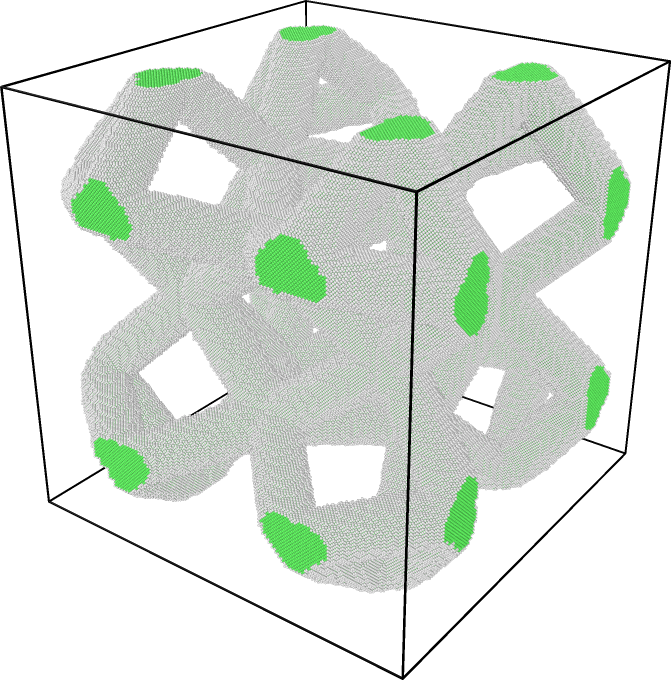}
    \caption{}
    \label{}
  \end{subfigure}
  \hfill
  \begin{subfigure}[b]{0.24\linewidth}
    \centering
    \includegraphics[width=\linewidth]{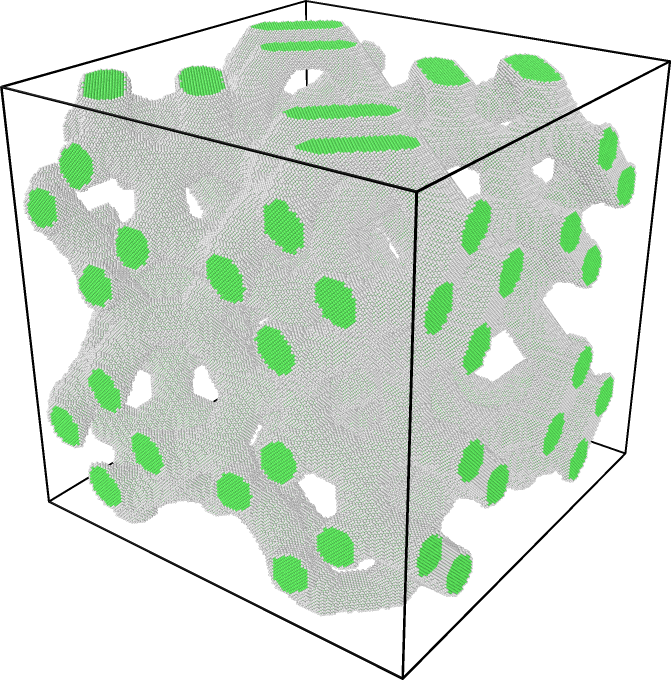}
    \caption{}
    \label{}
  \end{subfigure}
  \caption{Examples of topologically defined structures.}
  \label{fig:topologies}
\end{figure}

We tested the best-performing model (trained only on stochastic nanoporous structures) on a separate set of 22 topologies from the RCSR (Reticular Chemistry Structure Resource) database. Each of the topologies was generated with five values of target solid fraction (0.1, 0.2, 0.3, 0.4, 0.5 - recalculated later for as-built solid fraction), resulting in a total of 110 structures; examples are visualized in Figure \ref{fig:topologies}. This set was labeled with MD-computed elastic constants, following the same procedure as the stochastic samples.

Following the same procedure as with stochastic structures, topologically defined samples were converted to a binary array of 80$\times$80$\times$80 voxels to obtain CNN-predicted values of elastic constants. When the model was directly applied to architected structures with no fine-tuning, prediction errors increased substantially, compared to the model's performance on the stochastic test set. This behavior is expected, as the model was never trained on architected structures and cannot accurately extrapolate to unseen data. Despite the large prediction errors, the model preserved the relative ordering of stiffness across the new dataset. The Spearman rank correlation between MD-calculated and CNN-predicted elastic constants is $\rho$ = 0.955 and p-value = 3.26$\times10^{-58}$, indicating a strong monotonic agreement. In practice, it means that the model is able to identify which topologies have higher or lower elasticity compared to others, even though its absolute values of predictions are biased. 

\begin{figure}
    \centering
    \includegraphics[width=0.7\linewidth]{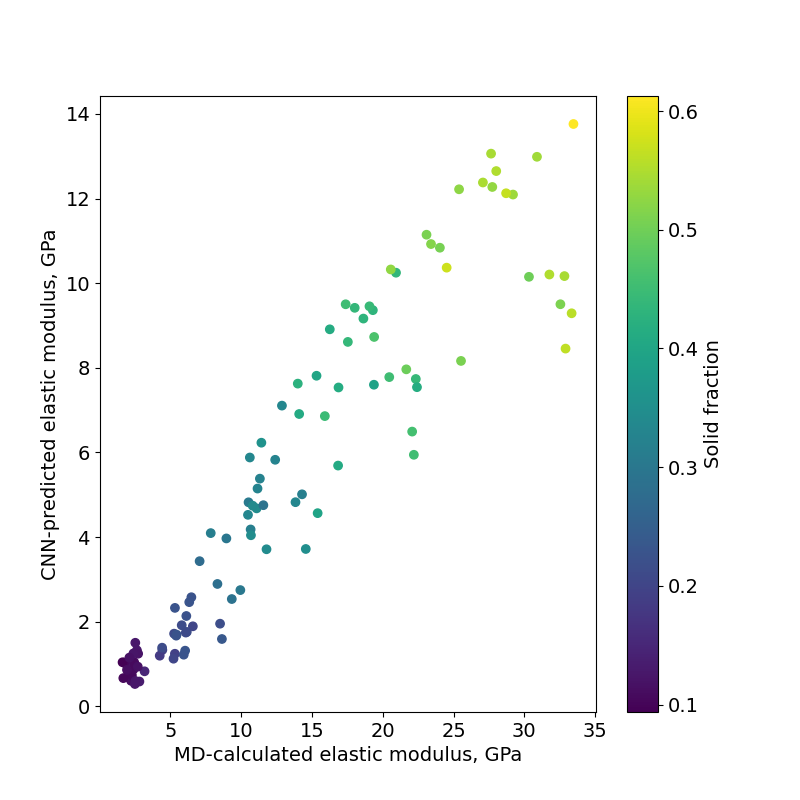}
    \caption{MD-calculated vs CNN-predicted elastic constants for dataset of RCSR topologies.}
    \label{fig:topo_scatter}
\end{figure}

\section{Hybrid model architecture}
\label{si:hybrid_model}
This architecture is a hybrid neural network model that fuses convolutional and dense feature representations to enable joint learning from binary vectorized structural inputs and handcrafted descriptor-based features. It is designed to perform a regression task where both spatial and tabular data provide complementary information. 

\begin{figure}
    \centering
    \includegraphics[width=0.7\linewidth]{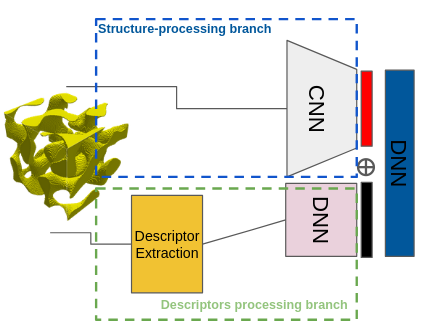}
    \caption{Hybrid model architecture.}
    \label{fig:si_hybrid_model}
\end{figure}

The first branch uses a convolutional neural network to extract high-level features from structured input data. This part of the model processes the data through a convolutional backbone (such as DenseNet-201) to produce a feature vector (denoted in red in the Figure \ref{fig:si_hybrid_model}). These extracted features represent complex spatial or hierarchical patterns present in the input. In the case of DenseNet-201, this feature vector has a dimension of 768.

The second branch handles a separate input consisting of numerical descriptors as listed in Table 1 of the main manuscript. This input is passed through a sequence of linear transformations, activation functions, and dropout layers (depicted by pink rectangle in the Figure \ref{fig:si_hybrid_model}) designed to distill the information into a compact, informative representation. This fully-connected network consists of 3 linear layers, each having 33 nodes, connected with ReLU activation function and drop-out layers with dropout probability of 0.3 and produces a feature vector (denoted as black rectangle) of dimension 33.
Once both branches have processed their respective inputs, their outputs — one from the CNN and one from the fully connected network — are concatenated into a single combined feature vector. This combined representation is then passed through another sequence of fully connected layers (represented as blue DNN network) that further integrate the information and transform it down to a final prediction, typically a single output value.

\bibliography{sn-bibliography}

\end{document}